\documentclass[prb,aps,amsmath,amssymb,nofootinbib,epsfig,showpacs,twocolumn,superscriptaddress]{revtex4-1}

\usepackage{amsmath,amssymb,amsfonts,graphicx,multirow,color,bbm,textcomp,mathtools,dsfont}
\usepackage{paralist}
\usepackage{srcltx}
\usepackage[all,cmtip]{xy}
\usepackage{mathrsfs}
\usepackage{srcltx,soul,subfigure}
\usepackage{txfonts}

\newcommand{\bra}[1]{\langle #1 |}
\newcommand{\ket}[1]{| #1 \rangle}
\newcommand{\braket}[2]{\left \langle #1 | #2 \right\rangle}

\newcommand{\bee}{\begin{equation}}
\newcommand{\ee}{\end{equation}}
\newcommand{\bma}{\begin{pmatrix}}
\newcommand{\ema}{\end{pmatrix}}
\newcommand{\balig}{\begin{align}}
\newcommand{\ealig}{\end{align}}

\newcommand{\bZ}{\mathbb{Z}}
\newcommand{\ba}{\begin{align}}
\newcommand{\ea}{\end{align}}

\newcommand{\ignore}[1]{}

\usepackage{siunitx}

\newcommand{\bI}{\mathbbm{1}}

\newcommand{\bk}{{\bf k}}

\usepackage{dcolumn}
\newcolumntype{C}[1]{>{\centering\let\newline\\\arraybackslash\hspace{0pt}}m{#1}}

\begin{document}


\title{Quantized Berry Phase and Surface States under Reflection Symmetry or Space-Time Inversion Symmetry}

\author{C.-K. Chiu}
\affiliation{Condensed Matter Theory Center and Joint Quantum Institute and Maryland Q Station, Department of Physics, University of Maryland, College Park, MD 20742, USA}
\affiliation{Kavli Institute for Theoretical Sciences, University of Chinese Academy of Sciences, Beijing 100190, China}
\author{Y.-H. Chan}
\affiliation{Institute of Atomic and Molecular Sciences, Academia Sinica, Taipei 10617, Taiwan}

\author{A. P. Schnyder}
\affiliation{Max-Planck-Institut f\"ur Festk\"orperforschung, Heisenbergstrasse 1, D-70569 Stuttgart, Germany}

%

\date{today}

\begin{abstract}
 As reflection symmetry or space-time inversion symmetry is preserved, with a non-contractible integral loop respecting the symmetry in the Brilliouin zone, Berry phase is quantized in proper basis. Topological nodal lines can be enclosed in the integral loop and $\pi$-Berry phase topologically protects the nodal lines. In this work, we show that to have quantized Berry phase restricted by the symmetry in any crystal structure, we choose to use the cell-periodic convention  and define the origin point in the real space at one of the reflection (inversion) centers. In addition, $\pi$-Berry phase is not the sufficient condition leading to the presence of the stable surface states. Their presence crucially depends on the location of the termination and the crystal structure in the unit cell. By using these new conditions we further reexamine if stable surface states exist in the known topological nodal line materials stemming from reflection symmetry or space-time inversion symmetry. 
\end{abstract}

\maketitle

\section{Introduction}

	The topological states of matter in solid state systems have attracted scholarly attentions in the physics community. The key signature in most of the topological states is the presence of the stable boundary states\cite{changechern,Hatsugai93}, which are robust against symmetry-preserving disorders. Under symmetry operations, the entire topological systems are always invariant. A topological invariant, which is quantized by the symmetries, characterizes a topological system and determines the number of the stable boundary states\cite{Thouless:1982rz,PhysRevLett.95.226801,Kane:2005vn}. In particular, the robustness of the boundary states leads to universal physical observables, such as quantized Hall conductance\cite{Klitzing}. One of the prominent examples is Majorana bound states protected by particle-hole symmetry\cite{Kitaev2001,Zhang:2018aa,Wangeaao1797}; the non-zero $Z_2$ invariant defined in the 1D superconductor indicates the appearance of the Majorana bound state.

\begin{figure*}[t!]
\begin{center}
\includegraphics[clip,width=1.55\columnwidth]{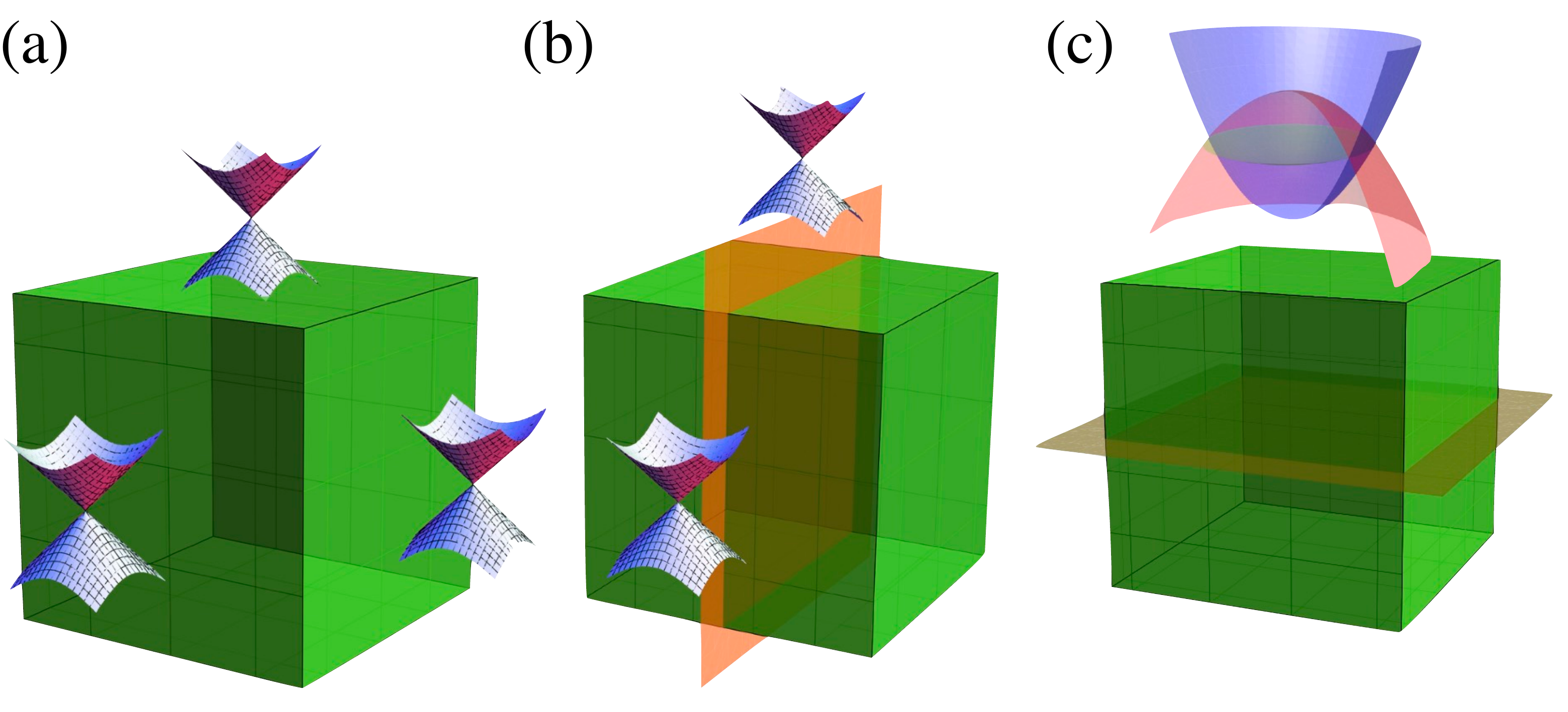}
  \caption{Stable surface states are present on the surfaces of the topological systems. The systems (b,c) preserve reflection symmetry with the orange reflection planes. (a) the topological insulators and superconductors in the ten-fold classification possess gapless surface states protected by symmetries at any termination on any surface. The key reason for the protection is that each surface is invariant under the non-spatial symmetry operations. (b) the stable gapless surface states appear at any termination on the surface preserving the reflection symmetry. (c) the drumhead surface state connecting the topological nodal ring appears on the surface, which does not preserve the reflection symmetry individually. The presence of this surface state depends on the details of the termination. }  
  \label{surface_states_all}
  \end{center}
\end{figure*}

	In the ten-fold classification of topological insulators and superconductors\cite{Kitaev2009,SchnyderAIP,Ryu2010ten,Schnyder2008,chiu_RMP_16}, the non-zero strong indices of the topological invariants indicate that stable surface states are present at \emph{any} termination on \emph{any} surface as illustrated in Fig.~\ref{surface_states_all}(a). Since time reversal symmetry, particle-hole symmetry, and chiral symmetry in the ten-fold classification are non-spatial, any termination alone on any surface always preserves the non-spatial symmetries, which protect the surface states. Furthermore, for topological crystalline insulators and superconductors\cite{Hsieh:2012fk,Tanaka:2012fk,Dziawa:2012uq,Xu:2012,Chiu_reflection,Sato_Crystalline_PRB14}, the crystalline symmetries are spatial so that the stable surface states can be present at any termination on the surface preserving crystalline symmetries as illustrated in Fig.~\ref{surface_states_all}(b). In these topological systems, surface states are always present on the surface individually preserving the symmetries. This is known as the standard bulk-boundary correspondence in topological systems.

	It has been experimentally observed that drumhead surface states appear\cite{PhysRevB.96.161112,Bian:2016aa} on the surfaces of topological nodal line semimetals. The bulk nodal lines are protected by reflection symmetry or space-time inversion symmetry (the combination symmetry of time-reversal and space inversion)\cite{ChiuSchnyder14,nodal_line_Yang,nodal_line_Yamakage,doi:10.1080/23746149.2017.1414631}. The surface hosting the drumhead surface state alone is not invariant under the symmetry operation as illustrated in Fig.~\ref{surface_states_all}(c). The drumhead surface state is not protected by the symmetries, which are preserved in the entire system. 
	This is another type of topological crystalline surface states. 
	The problem can be simplified to the 1D wire problem that the bulk is invariant under reflection (space-time inversion) operation and one wire end is mapped to the other end. The Berry phase as a topological invariant is quantized when the symmetry operators are momentum-independent\cite{Hughes:2011uq}. Although it has been shown that $\pi$-Berry phase leads to the presence of the symmetry-protected \emph{end} state and the termination of the surface must include a reflection (inversion) center\cite{PhysRevB.95.035421}, the criterion determining the presence of the stable end state in these systems has not been studied throughly. Furthermore, once the symmetry operators are momentum-dependent, unfortunately the Berry phase cannot be quantized in general and the stable end state might be absent.  



In this manuscript, we consider the relation between the Berry phase and the stable end state in the presence of reflection symmetry and space-time inversion symmetry respectively. Space-time inversion symmetry indicates that the system is invariant under the combined operations of time-reversal and space inversion and its symmetry operator is labelled by $TP$. In the following, we always consider $TP^2=\bI$ corresponding to spinless systems since only space-time inversion symmetry with $TP^2=-\bI$ cannot lead to protected nodal lines~\footnote{Ref.~\onlinecite{Sato_Crystalline_PRB14} implicitly shows the topological classification of 1d space-time inversion symmetric insulators. For $TP^2=\bI$, $K^A_{\mathbb{C}}(s=0;d=1,d_\parallel=1,0,0 )=\pi_0(\mathcal{R}_1)=\bZ_2$ and for $TP^2=-\bI$, $K^A_{\mathbb{C}}(s=4;d=1,d_\parallel=1,0,0 )=\pi_0(\mathcal{R}_5)=0$.  }.  


%


To compute the Berry phase, the bloch wavefunctions are usually written in the two conventions (unit-cell and cell-periodic) to compute topological invariants. The reason to choose the unit-cell convention is that all topological invariants in the ten-fold classification of topological insulators and superconductors\cite{Ryu2010ten,chiu_RMP_16,review_TIb} are defined in this convention and are always gauge invariants. Therefore, we examine the Berry phase in this convention, which is called the intercellular Zak phase\cite{PhysRevB.95.035421}.  In the cell-periodic convention the Berry phase as the Zak phase\cite{PhysRevLett.62.2747} corresponds to electronic polarization~\cite{kingSmithPRB93a,kingSmithPRB93b} in the 1D system with the terminations. Furthermore, the value of this Zak phase depends on the choice of the spatial origin point\cite{PhysRevB.95.035421,Vanderbilt_berry}. However, it has not been discussed before that the Berry phases quantized by the symmetry relies on the specific choice of the convention and the origin point; the fundamental reason that $\pi$-Berry phase leads to the presence of the surface state is missing in the literature.

To simplify the problems of the quantized Berry phase and the surface state, we focus on only 1D tight-binding systems at the beginning since the physics can be easily extended to higher dimensions. The Hamiltonians in the unit-cell convention are defined based on the hopping terms written in the distance between the different unit cells and is invariant under momentum shift $k \rightarrow k+2\pi$. The hopping terms in the cell-periodic convention is based on the physical distance of the atoms so that $H(k+2\pi)=H(k)$ does \emph{not} hold in general. Hence, the wavefunction can be written in the different conventions: the unit-cell convention $\ket{\phi_{k,j}}$ and the cell-periodic convention (atom-distance) $\ket{u_{k,j}}$. The relation between the two conventions is given by 
\bee
U_{\beta \alpha}(k)\ket{\phi_{k,j}}_\alpha = \ket{u_{k,j}}_\beta, \label{two conventions}
\ee
where label $j$ indicates the wavefunctions at the different energy levels $E_j$, labels $\alpha,\ \beta$ represent generic atom and orbital indices, $U$ is in the form of a diagonal unitary matrix $U(k)=\rm{diag}$$(e^{-ik r_1},e^{-ik r_2},\ldots )$, $r_\alpha$ is the location of the $\alpha$-th atom in unit cell. We remove $r$-dependence in the Bloch wavefunction since in the tight-binding model each orbital is located at \emph{one} point and this simplification does not alter the fundamental physics of the surface states discussed in this manuscript. The wavefunction $\ket{u_{k,j}}$ is known as the cell-periodic functions and the wavefunction in the unit-cell convention obeys $\ket{\phi_{k,j}}=\ket{\phi_{k+2\pi,j}}$. We note that the choice of the origin point is important since the unitary matrix $U$ can be changed by redefining the locations of the atoms.

In the unit-cell convention, when reflection (inversion) center is not located at the unit cell center, the operators of these two symmetries might be momentum-dependent in the lattice model. In the manuscript, we show that once the symmetry operators are momentum-dependent, the Berry phase can not be guaranteed to be quantized. In the cell-periodic convention, the Berry phase can be quantized when reflection (inversion) center is located at the spatial origin point. Furthermore, although it was thought that $\pi$ Berry phase leads to the presence of the surface state, the surface state might be even absent when the termination does not include the reflection (inversion) center.

The remainder of the manuscript is organized as follows. In Sec.~\ref{Berry sec} in generic 1d reflection symmetric systems, we show that the Berry phase in the unit-cell convention cannot be quantized and in the cell-periodic convention can be quantized when the reflection is at the spatial origin. The criterion of the stable surface state is throughly discussed in Sec.~\ref{surface_state}. In Sec.~\ref{CaAgAs}, CaAgAs is the example failing the surface state criterion of the reflection symmetry but possessing the surface state protected by other symmetries. In Sec.~\ref{time-reversal inversion section}, for space-time inversion symmetry we use the similar analysis to discuss the Berry phase and the surface state. We provide the example of the space-time inversion symmetric system (CaP$_3$) to show that the appearance of the stable surface state stems from other symmetries in Sec.~\ref{CaP3}, when the surface state criterion of the space-time symmetry is not obeyed. In Sec.~\ref{conclusion}, we summarize the criterion of the stable surface state in the reflection (space-time inversion) symmetric systems. Some technical details are relegated to Appendix.



\section{Berry phase with reflection symmetry} \label{Berry sec}


To study the relation between the Berry phase and the surface state in reflection symmetric systems, we consider a 1D chain with lattice constant $a=1$ and examine the Berry phases in these different conventions.  

\begin{figure}[t!]
\begin{center}
\includegraphics[clip,width=0.85\columnwidth]{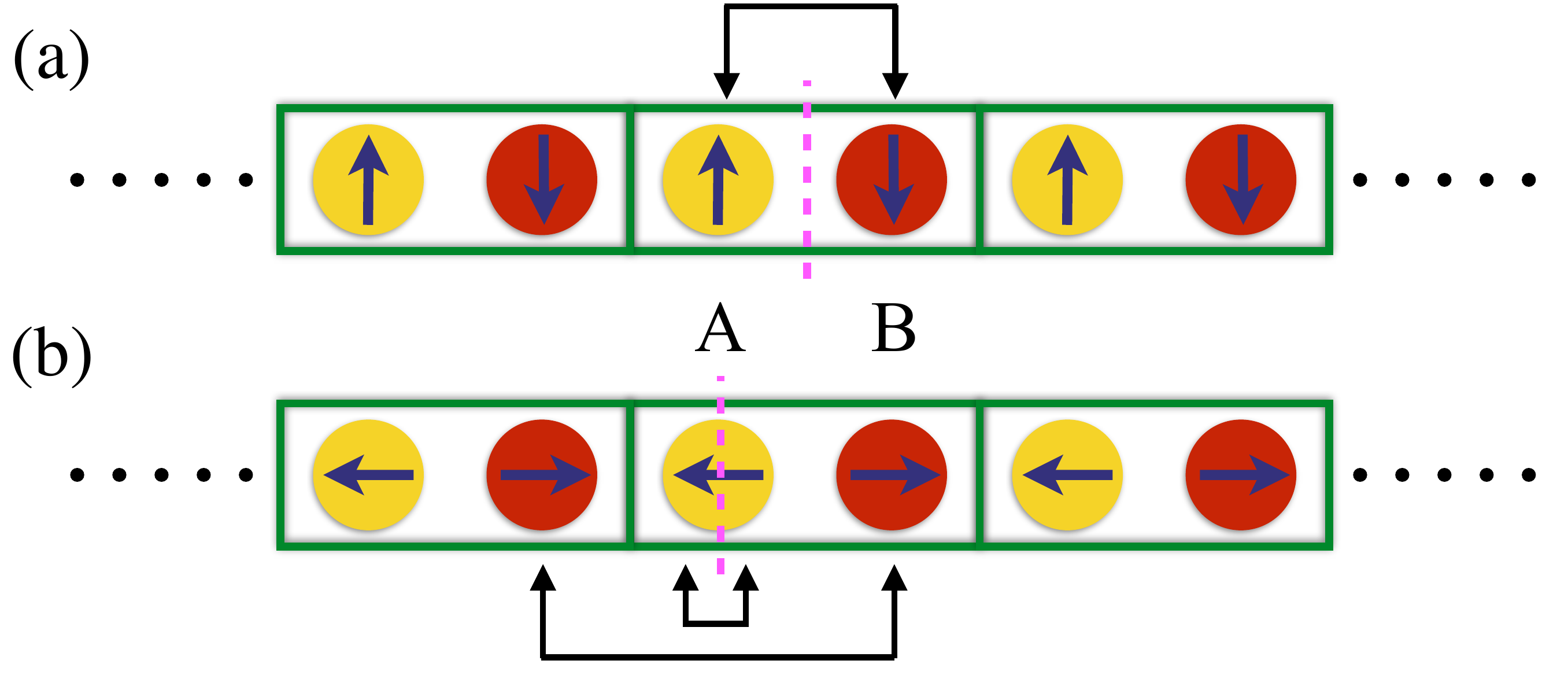}
  \caption{ The pink dashed lines represent the reflection center and the green boxes represent the unit cells. The reflection operator in the $x$ direction is given by $R_x=\sigma_x$ ($i$ is neglected) for the spin-$1/2$ systems.  (a) the reflection center is located at the center of the unit cell. All of the atoms in one unit cell are reflected to the same unit cell and spin up and down flip under the reflection operation. (b) while the reflection of the yellow atom is itself in the same unit cell, the red atom is reflected to another unit cell. This discrepancy leads to $k$-dependent reflection operator. The reflection operation does not change the spins in the $x$ direction. } 
  \label{k_dependent}
  \end{center}
\end{figure}

We first consider a 1D model preserving reflection symmetry in the unit-cell convention so the Hamiltonian in the momentum space obeys 
\bee
R_{-k}H(-k)R_k=H(k), \label{reflection eq}
\ee
where the reflection operator obeys $R_{k}^\dagger=R_{-k}$ and $R_{-k}R_{k}=\bI$. The momentum label $k$ in $R_k$ indicates the reflection symmetry operator might be momentum-dependent. By assuming the absence of energy degeneracies, the reflection symmetry acts the wavefunctions $\ket{\phi_{k,j}}$ in the unique expression
\bee
\ket{\phi_{-k,j}}=e^{-i\alpha_k^j}R_k\ket{\phi_{k,j}}.
\ee
The reflection phase $\alpha_k^j$ connects the wavefunction at $\pm k$. It has been shown that as $R_k$ is momentum-independent, the Berry phase is quantized to 0 or $\pi$~[\onlinecite{Hughes:2011uq}]. Now we carefully consider the Berry phase with the momentum-dependent reflection operator. The integral path of the Berry phase has to be invariant under reflection to \emph{potentially} quantize the Berry phase. In this 1D model, we simply choose the integration path as the entire 1D BZ and write the Berry phase (the intracellular Zak phase in [\onlinecite{PhysRevB.95.035421}]) in the integral form
\begin{small}
\begin{eqnarray}
\mathcal{P}&=&-i \big ( \int_0^{\pi} + \int_{-\pi}^0 \big) \sum_{E_j< E_{\textrm{F}}}      \bra{\phi_{k,j}} \partial_{k} \ket{\phi_{k,j}} dk           \nonumber \\
&=&-i \int_{0}^{\pi}  \sum_{E_j< E_{\textrm{F}}}      \bra{\phi_{k,j}} \partial_{k} \ket{\phi_{k,j}} dk           \nonumber \\
 &\ &+i \int_{0}^{\pi}  \sum_{E_j< E_{\textrm{F}}}      \bra{\phi_{k,j}} R_{k}^\dagger e^{i\alpha_k^j} \partial_{k} e^{-i\alpha_k^j } R_{k}\ket{\phi_{k,j}} dk \nonumber \\
&=&\sum_{E_j  < E_{\textrm{F}}} (\alpha_\pi^j - \alpha_{0}^j) +  i \int_{0}^{\pi}  \sum_{E_j< E_{\textrm{F}}}      \bra{\phi_{k,j}} R_{k}^\dagger ( \partial_{k} R_{k} ) \ket{\phi_{k,j}} dk  \ (\rm{mod}\ 2\pi), \label{Berryphase}
\end{eqnarray}
\end{small}
where $E_F$ is the Fermi level. We have $R_0^2=R_\pi^2=\bI$ \footnote{for spin-$1/2$ systems, $i$ in $R_k$ is always dropped.} so $\alpha_0^j,\ \alpha_\pi^j$ are either $0$ or $\pi$; the first term is quantized and given by 
\bee
\sum_{E_j  < E_{\textrm{F}}} (\alpha_\pi^j - \alpha_{0}^j )=0,\ \pi \quad (\rm{mod}\ 2\pi),
\ee
where $n$ is an integer. The second term in Eq.\ \ref{Berryphase} vanishes when the reflection operator $R_k$ is momentum-independent. That is, physically when the reflection image of each atom in the unit cell is in the same unit cell as shown in Fig.~\ref{k_dependent}(a), the reflection center is at the unit cell center and $R_k$ does not depend on $k$. 
Hence, the Berry phase under reflection symmetry is either $0$ or $\pi$, 
since $2\pi$ phase can be removed by large $U(1)$ gauge transformation. Furthermore, in the unit-cell convention $H_{k=\pi}=H_{k=-\pi}$ so that at the high symmetry points ($k_0=0,\ \pi$) the reflection symmetry operator $R_{k_0}$ commutes with the Hamiltonian $H_{k_0}$; hence, we can further define $n_{0}^+$ and $n_{\pi}^+$ as the number of the occupied states in the $R_{k_0}=1$ eigenspace at $k_0=0,\ \pi$ respectively in the form of 
\bee
n_{k_0}^+=\sum_{E_j < E_F} \bra{\phi_{k_0,j}} \frac{\bI+R_{k_0}}{2}\ket{\phi_{k_0,j}} \label{occupation number}
\ee
We note that in the cell-periodic convention $H_{k=\pi}=H_{k=-\pi}$ does \emph{not} hold so that $n_{\pi}^+$ at $k=\pi$ is ill-defined. The reflection occupation numbers are directly related to the reflection phases
\bee
(-1)^{n_0^++n_\pi^+}=e^{i\sum_{E_j<E_F}(\alpha_\pi^j-\alpha_0^j)}.
\ee 
Hence, $\pi$ Berry phase corresponds to odd number of $n_0^++n_\pi^+$. We note that due to momentum-independent $R_k$, $(-1)^{n_0^++n_\pi^+}=(-1)^{n_0^-+n_\pi^-}$, where $n_{0}^-$ and $n_{\pi}^-$ is the number of the occupied states in the $R_{k_0}=-1$ eigenspace at $k_0=0,\ \pi$.

However, as $R_k$ is momentum-dependent, the reflection symmetry operator has a generic block-diagonalized form 
\bee
R_k=\phi_{i_1j_1}e^{in_1 k }\oplus \phi_{i_2j_2}e^{in_2 k }\oplus\ldots \oplus \phi_{i_Nj_N}e^{in_N k },
\ee
where $\phi_{i_lj_l}$ is a unitary matrix obeying $\sum_{j_l}\phi_{i_lj_l}\phi_{j_lk_l}=\delta_{i_lk_l}$ (due to $R_{-k}R_{k}=\bI$), $n_j$ indicates atoms in the $(n_j+i)$-th unit cell are reflected to atoms in the $(-i)$-th unit cell as shown in Fig.~\ref{k_dependent}(b), and we use the lattice constant $a\equiv 1$. We have 
\bee
R_k^\dagger \partial_k R_k= in_1\delta_{i_1j_1}\oplus in_2\delta_{i_2j_2} \oplus \ldots \oplus in_N\delta_{i_Nj_N}.
\ee
Unfortunately, generic state $\ket{\phi_{k,j}}$ might mix at least two different blocks in $R_k$ and preserves the reflection symmetry. Non-quantized $i \int_{0}^{\pi}  \sum_{E_j< E_{\textrm{F}}}      \bra{\phi_{k,j}} R_{k}^\dagger (\partial_{k} R_{k} )\ket{\phi_{k,j}} dk$ leads to unquantized Berry phase (see the example in Appendix \ref{unBerry}). Since it has been mistakenly showed in Ref.~\onlinecite{nodal_line_Yang,nodal_line_Yamakage} that this term can be quantized as $R_k$
 is k-dependent, it is important to show the Berry phase is not always quantized in the unit-cell convention and the unquantized quantity cannot be used to study the presence of the surface state. Furthermore, the odd occupation number, which cannot lead to $\pi$ Berry phase, is not qualified to determine the topology in the 1d chain.


Next we study the Berry phase in the cell-periodic convention. In this convention, the reflection symmetry operator can be decomposed to $k$-independent matrix with a global $k$-dependent phase 
\bee
R'_k=e^{2ik d_m} R_o, \label{R cell-periodic}
\ee 
where $R_o$ is a $k$-independent matrix and $d_m$ is the distance between the origin point and the location of the reflection center (see Appendix \ref{Ratomdistance} for the proof). Here, we define the origin point is located at one of the centers of the unit cells. Similarly, the wavefunctions at $\pm k$ has the same relation stemming from reflection symmetry
\bee 
\ket{u_{-k,j}}=e^{-i\alpha_k^j}R_k'\ket{u_{k,j}}
\ee
The Berry phase in this convention is given by 

\begin{small}
\begin{eqnarray}
\mathcal{P}'&=&-i  \int_{-\pi}^\pi \sum_{E_j< E_{\textrm{F}}}      \bra{u_{k,j}} \partial_{k} \ket{u_{k,j}} dk           \nonumber \\
&=
&\sum_{E_j  < E_{\textrm{F}}} (\alpha_\pi^j - \alpha_{0}^j) +  i \int_{0}^{\pi}  \sum_{E_j< E_{\textrm{F}}}      \bra{u_{k,j}} R_{k}^\dagger ( \partial_{k} R_{k} ) \ket{u_{k,j}} dk \nonumber \\
&=
& n\pi - 2d_m n_o \pi\ (\rm{mod}\ 2\pi),
 \label{Berryphase II}
\end{eqnarray}
\end{small}
where $n_o$ is the number of the occupied bands (see an example in Appendix \ref{Berry_atom}). Hence, $\mathcal{P}'$ is quantized $(0,\pi)$ as the reflection center is located at the unit cell center and boundary ($d_m=0,\ 0.5a$). The two different values of the Berry phase indicates the two distinct topological phases. The polarization of the occupied Wannier functions has this relation with the Berry phase~\cite{PhysRevLett.62.2747,doi:10.1063/1.2743018}
\bee
\frac{a}{2\pi}\mathcal{P}'=\sum_{E_j<E_F}\int_{\Omega} x |\phi_j(x)|^2 dx =\bar{X}\ \mathrm{mod}\ a, \label{center_W}
\ee
where $\Omega$ is the region of the single unit cell. 
Since the origin point is chosen at the unit cell center, the center of the wavefunctions are located at the unit cell center as $\mathcal{P}'=0$. 
On the other hand, as $\mathcal{P}'=\pi$, the center of the wavefunctions is located at the unit cell boundary, which is a half lattice constant away from the origin point. 



%

\subsection{Surface states} \label{surface_state}

In the literature, it has been shown that stable surface states appear as Berry phase is $\pi$. However, to have the stable surface states, the location of the termination plays an important role to determine the presence of the stable surface states. 

\begin{figure}[t!]
\begin{center}
\includegraphics[clip,width=0.75\columnwidth]{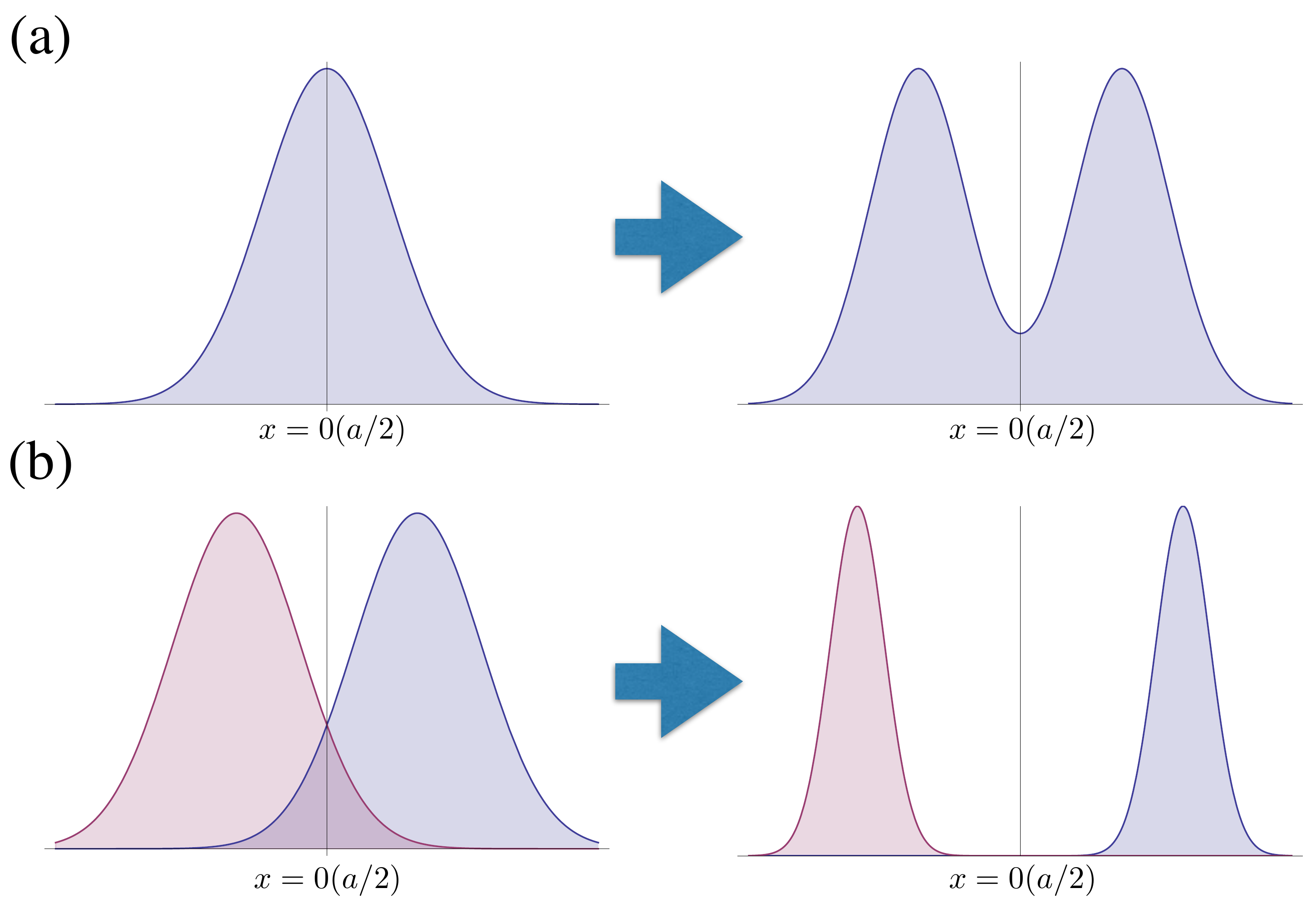}
  \caption{ The Wannier function distribution in the reflection symmetric wire. (a) the center of a single wavefunction is located at one of the reflection centers ($x=0,\ a/2$) with the Berry phase $\mathcal{P}'=0,\ \pi$. The $\pi$ Berry phase indicates the center of the single wavefunction is locked at $x=a/2$. The termination at $x=a/2$, which cuts the wavefunctions to two halves, leads to the presence of the stable surface state. (b) the center of two wavefunctions is located at one of the reflection centers and the Berry phase $\mathcal{P}'=0$. The terminations at the reflection center might not cut any wavefunction since the center of each wavefunction can move freely with respect to reflection symmetry. Therefore, the surface states might be absent.    } 
  \label{wavefunction}
  \end{center}
\end{figure}

\begin{figure}[t!]
\begin{center}
\includegraphics[clip,width=0.95\columnwidth]{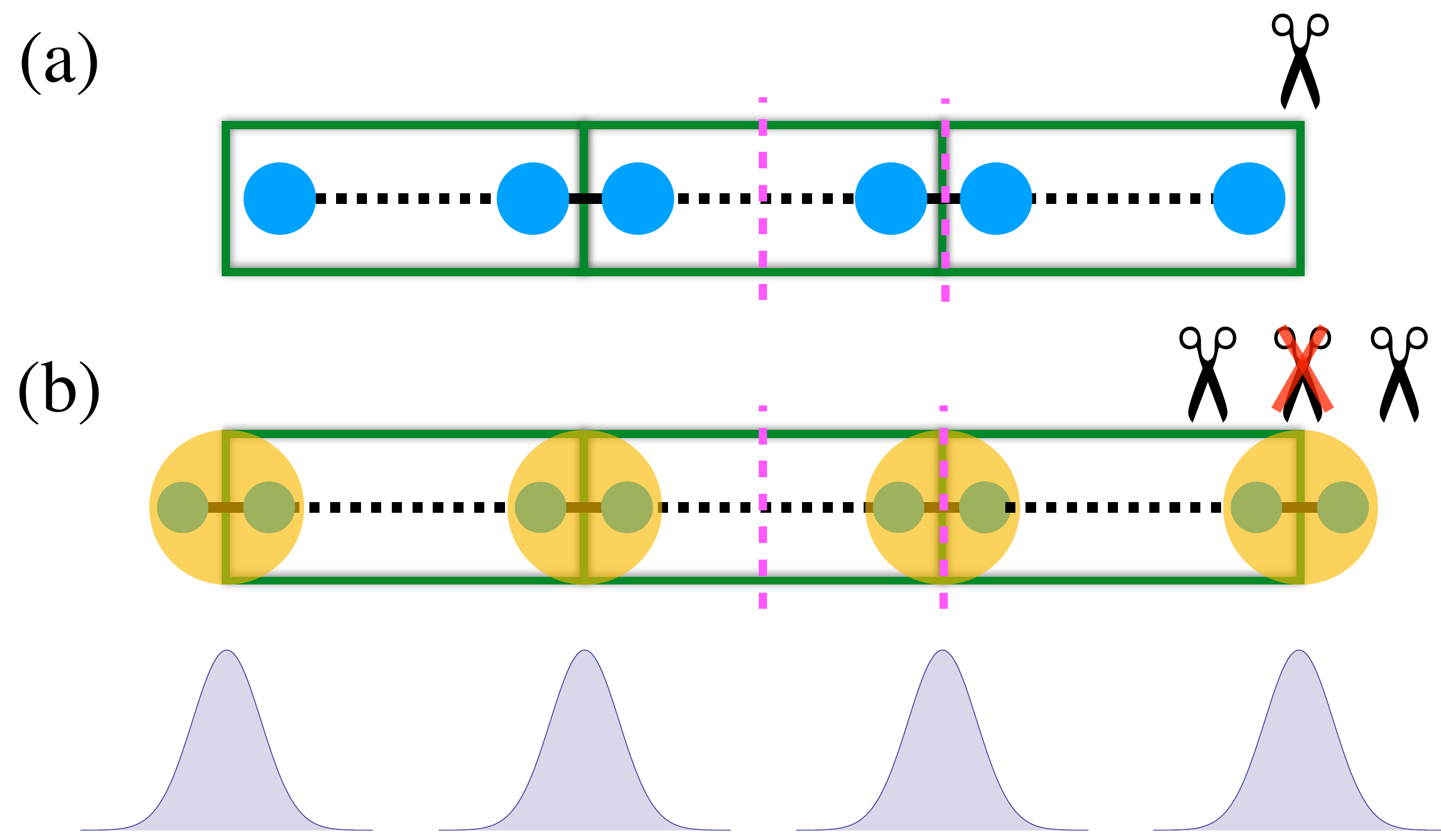}
  \caption{ The solid (dashed) lines represent the strong (weak) bonds and the pink lines indicate the reflection centers. The Wannier function is located at the boundary of the unit cell. (a) the strong bond, which connects the two atoms, can be cut to form a dangling end state. (b) the strong bond is inside the orange atom sitting at the reflection center. Since it is impossible to separate the single atom, the strong bond cannot be cut to form the dangling end state.     } 
  \label{cut_reflection_center}
  \end{center}
\end{figure}

A straightforward way to understand the presence of surface states is from the center of the occupied Wannier functions. We choose the origin point to be a reflection center and the unit cell center $(d_m=0)$ so that the boundary of the unit cell ($x=a/2$) is also another reflection center. There are only four possible locations of the Wannier functions preserving reflection symmetry. (i) when the center of single or multiple Wannier functions is located at the origin point ($x=0$) as shown in Fig.~\ref{wavefunction}(a), the Berry phase in the cell-periodic convention vanishes. (ii) when the center of a Wannier function is off reflection center, due to the reflection symmetry, another Wannier function, which is the reflection of the original wavefunction, is located at the other side of the reflection center. The Berry phase of these wavefunctions should be zero (mod $2\pi$) since the center of the two wavefunctions is at either the origin point or $x=a/2$ as shown in Fig.~\ref{wavefunction}(b) ($\mathcal{P}'=0,\ 2\pi$). (iii) when two Wannier functions are located at the unit cell boundary, $2\pi$ Berry phase $\mathcal{P}'$, which is equivalent to zero under the gauge transformation, indicates the wavefunction can be smoothly deformed to two wavefunctions off reflection center without breaking reflection symmetry as demonstrated in Fig.~\ref{wavefunction}(b). Therefore, no locked surface states are present in any specific termination.  
(iv) when the center of a single Wannier function is located at the boundary of the unit cell ($x=a/2$) (Fig.~\ref{wavefunction}(a)), the Berry phase $\mathcal{P}'$ is given by $\pi$. If the center of the Wannier function moves away from the unit cell boundary, in the absence of another wavefunction as the reflection image, the reflection symmetry is broken. Hence, the center of the single Wannier function has to be \emph{locked} at the boundary of the unit cell by reflection symmetry. When the unit cell boundary is chosen as the termination, this Wannier function is always cut to halves to form a stable surface state, which is similar with the presence of a Majorana bound state in the Kitaev chain~\cite{Kitaev2006}, as illustrated in Fig.~\ref{cut_reflection_center}(a).


If the termination is at the origin point, which is one of the reflection cetners, although zero-value Berry phase indicates the center of Wannier functions located at the origin, the Wannier functions can be either locked or unlocked at the origin, since in case (ii) the Wannier functions are away from the origin. Therefore, the Berry phase in this basis cannot determine the presence of the stable surface state at this termination including the origin point. 
To determine the presence of the stable surface state we can simply exchange the definitions of the unit cell boundary and the origin point and then compute the new Berry phase in the same convention. The new origin point, which was the old boundary of the unit cell, is still a reflection center so that the Berry phase $P'_{0}$ is quantized. Similarly, $\pi$ Berry phase $P'_{0}$ indicates the presence of a surface state at the termination of the old origin point. Interestingly, the number ($n_0$) of the occupied surface states indicates the relation of the surface states at these two different reflection center terminations by eq.~\ref{Berryphase II}
\bee
(-1)^{n_0}=e^{i (P'_0+ P'_{a/2})},
\ee
where $P'_{a/2}$ is the Berry phase based on the previous origin point. As $n_0$ is even, the surface states appear or vanish at the two different reflection centers at the same time. As $n_0$ is odd, only one reflection center as a termination can host a stable surface state. 

Thus, to have a stable surface state protected by reflection symmetry, the termination must be located at a reflection center. The example in Appendix \ref{no_surface} shows the absence of the surface states when the termination is not located at any reflection center. 

When the origin is located at the unit cell center, the reflection operator is momentum-independent, and the reflection center is at the origin $d_m=0$. The Berry phases in the two conventions are identical
\bee
\mathcal{P}=\sum_{E_j  < E_{\textrm{F}}} (\alpha_\pi^j - \alpha_{0}^j)=\mathcal{P}'
\ee
In this special case, the Berry phases in the two different conventions are either $0$ or $\pi$. When $\mathcal{P}=\mathcal{P}'=\pi$, a surface state protected by reflection symmetry appears at the boundary of the unit cell center, which is a reflection center. 


It is common in real materials that some atoms are exactly located at the reflection centers. Once the termination is made at the reflection center, those atoms are either completely removed or included in the sample so that the termination can never be located at any reflection center as illustrated in Fig.~\ref{cut_reflection_center}(b). As the center of the Wannier functions is in these atoms, the surface state might be absent in the termination since the atom cannot be cut into two halves. 

In another point of view, the atoms located at the boundary of the unit cell, which is the reflection center, always leads to the momentum-dependent reflection operator in the unit-cell convention. Since the center of the unit cell is the other reflection center, the reflection images of the atoms inside the unit cell are in the same unit cell as illustrated in fig.~\ref{k_dependent}(a). Therefore, the inside-atom part of the reflection operator is momentum-independent. On the other hand, the atoms located at the unit cell boundary must be reflected to another boundary of the unit cell. Therefore, this part of the reflection operator must be momentum-dependent. In the unit-cell convention, the Berry phase might be unquantized due to the atoms at the unit cell boundary. This unquantized Berry phase directly leads to the absence of the surface  state. 


	The following example shows that when atoms are located at the reflection centers, $\pi$ Berry phase in the cell-periodic convention does not lead to the presence of the surface states at any termination. Consider A and B atoms located at $x=0,\ a/2$ respectively in the unit cell as shown in Fig.~\ref{k_dependent}(b). The atom configuration leads to the reflection centers at the atoms. The reflection operator transformed from real space to momentum space is given by 
\begin{align}
\hat{R}=&\sum_j \big ( A^\dagger_{-j}A_j + B^\dagger_{-j-1} B_j\big ) \\
=&\sum_k (A_{-k}^\dagger \ B_{-k}^\dagger )
R_k
\bma 
A_{k} \\
B_{k}
\ema
\end{align}
The matrix form of the reflection operator always depends on momentum 
\bee
R_k=
\bma 
1 & 0 \\
0 & e^{-ik}
\ema. \label{k-dependent R}
\ee
We simply consider the system only has sublattice potential difference and same-atom hopping so that the Hamiltonian can be written as 
\bee
 H=(m+\cos k) \sigma_z 
 \ee 
preserving the reflection symmetry $R_{-k}H(-k)R_k= H(k)$, where $|m|>1$ to have the insulating phase. 
In the cell-periodic convention, since the origin point is the reflection center and the location of atom A, the expression of the occupied state is given by 
\bee
\ket{u_k}=\left \{
\begin{array}{cc}
(1,0)^T,\quad m<-1 \\
(0,e^{-ik/2})^T,\quad  m>1
\end{array}
\right. 
\ee
the Berry phase is quantized 
\bee
P'= \left \{
\begin{array}{cc}
0,\quad m<-1 \\
\pi,\quad  m>1
\end{array}
\ (\rm{mod}\ 2\pi) \right. 
\ee
The $\pi$ Berry phase indicates the single occupied state located at the boundary of the unit cell, which is the location of atom B. Unfortunately, at the boundary the physical termination cannot be cut the atom into two halves and all of the occupied states are extended states (due to $\cos k$) so the surface states always are absent. When in the crystal structure the atoms are located at a reflection center as a termination, stable surface states are absent even if the Berry phase in the cell-periodic convention is $\pi$. We note that $\mathcal{P}'=0,\ \pi$ correspond to the two different trivial topological phases. That is, the two phases can adiabatically connect to the two different atomic insulators by turning off the hopping terms ($\cos k$). 




\subsection{Examples: CaAgAs} \label{CaAgAs}

\begin{figure}[t!]
\begin{center}
\includegraphics[clip,width=0.95\columnwidth]{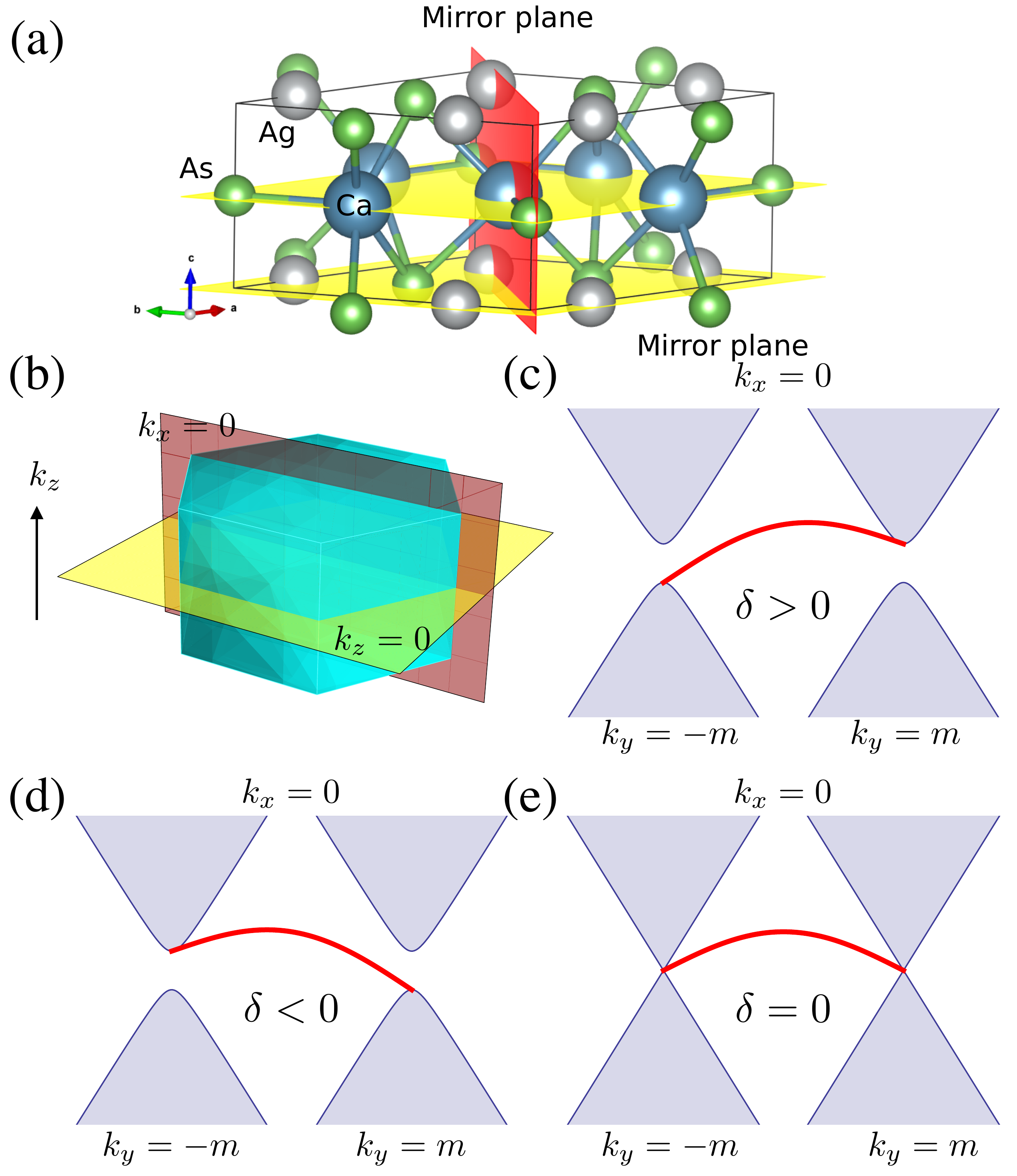}
  \caption{ (a) the crystal structure of CaAgAs shows the two mirror planes. The atoms are exactly located in the mirror planes in the $z$ direction; hence, the reflection symmetry in the $z$ direction cannot lead to a stable surface state. (b) The BZ of CaAgAs is a Hexagon prism and the two reflection planes in the two directions go through the BZ. (c-e) the energy dispersion on $(001)$ surface in the $k_x=0$ reflection plane. The presence of the spin-orbit coupling ($\delta$) destroys the nodal ring and leads to the non-zero mirror Chern number. The chiral surface mode (red) is always present and connect near the two points ($k_y=\pm m$) of the original nodal ring, when $\delta$ is either positive or negative as illustrated in panel (c,d). Hence, even without the spin-orbital coupling the surface state still appears and connects the nodal ring in panel (e). } 
  \label{CaAgAs_surface}
  \end{center}
\end{figure}

The semimetals CaAgP and CaAgAs have been studied to possess a stable 4-fold degenerate nodal ring protected by reflection symmetry along the $z$ direction in the absence of the spin-orbit coupling. Our focus is on CaAgAs to study the physics of the surface states. It has been shown in the DFT calculations\cite{nodal_line_Yamakage} and APRES experiment\cite{PhysRevB.96.161112} that a drumhead surface state emerges from the bulk nodal ring. It was thought that the presence of the drumhead surface state stems from the reflection symmetry. However, the centers of the reflection symmetry are exactly located at the atoms as shown in the crystal structure of Fig.~\ref{CaAgAs_surface}(a). Any of two terminations (Ca$_3$As and Ag$_3$As$_2$) in CaAgAs on $(001)$ surface is unable to host a stable surface state protected by the reflection symmetry in the $z$ direction since the termination cannot be at any reflection centers. Hence, the presence of the surface states does not stem from this reflection symmetry. In the following, we show non-zero mirror Chern numbers from the reflection symmetries perpendicular to the $z$ direction leads to the surface state.  

	The space group of CaAgAs ($\# 189$) has three non-identity symmetry generators $\bar{R}_z\vec{r}=(x,y,-z)$, $\bar{C}_3\vec{r}=(\frac{-x+\sqrt{3}y}{2},\frac{-y-\sqrt{3}x}{2},z)$, $\bar{R}_x\vec{r}=(-x,y,z)$ [\onlinecite{aroyo_bulg_2011}]. Since the 4-fold degenerate nodal ring is located $k_z=0$ plane, the low energy effective model can be written near $k_z=0$ 
\bee
H(k_\parallel,k_z)=(k_\parallel^2-m^2)\tau_x\otimes \sigma_0 + k_z \tau_y \otimes \sigma_0  \label{CaP_H},
\ee	
where $\tau_\alpha$ and $\sigma_\beta$ represent orbital and $1/2$-spin indices respectively, $k_\parallel$ is $(k_x,k_y)$, and $m$ is a positive constant. The nodal ring appears at $k_z=0$ and $|k_\parallel |=m$.  This effective Hamiltonian preserves time-reversal symmetry and the crystalline symmetries of these three generators above. That is, the Hamiltonian obeys $TH(-k_\parallel,-k_z)T^{-1}=H(k_\parallel,k_z)$ with time reversal symmetry operator $T=\tau_0\otimes\sigma_y \mathcal{K}$, where $\mathcal{K}$ is the complex conjugate operator. Likewise, the crystalline symmetries lead to $R_zH(k_\parallel,-k_z)R_z^{-1}=H(k_\parallel,k_z)$ with $R_z=\tau_x\otimes \sigma_z$, $C_3H(\bar{C}_3^{-1}k_\parallel,k_z)C_3^{-1}=H(k_\parallel,k_z)$ with $C_3=\tau_0\otimes e^{i\pi\sigma_z/3}$, $R_xH(-k_x,k_y,k_z)R_x^{-1}=H(k_\parallel,k_z)$ with $R_x=\tau_0\otimes \sigma_x$. Fig.~\ref{CaAgAs_surface}(b) shows two reflection planes along the $x$ and $z$ directions. We note that the identity matrix $\tau_0\otimes\sigma_0$, which does not alter the physics, is neglected although the identity matrix can be also added in the Hamiltonian without breaking any symmetry.

In the absence of the spin-orbital coupling, because spin $SU(2)$ symmetry is preserved, the system can be treated as two identical spinless systems $h(k_\parallel,k_z)=(k_\parallel^2-m^2)\tau_x+ k_z \tau_y $, each 2-fold degenerate nodal ring is protected by the reflection symmetry along the $z$ direction since the two bands of the ring correspond to the different reflection eigenvalues. However, this reflection symmetry cannot lead to the presence of the drumhead surface states due to the crystal structure (Fig.~\ref{CaAgAs_surface}(a)). 

	To examine the surface states, we introduce the spin orbital coupling, which breaks spin $SU(2)$ symmetry and preserves all of the other system symmetries. Hence, the spin-orbital coupling destroying the nodal ring can be written as
\bee
\delta \tau_z \otimes (-k_y \sigma_x + k_x \sigma_y ).
\ee 
We note that the spin-orbital coupling does not have the unique form. Under the symmetry constraints, the different forms of the spin-orbital coupling do not change the surface physics. Since the reflection symmetry in the $x$ direction is preserved, the mirror Chern number can be computed at $k_x=0$ mirror plane at one of the reflection eigenspaces. We consider the Hamiltonian projection in $k_x=0$ mirror plane at $R_x=1$ reflection eigenspace near the two lowest energy points $k_y\sim  m,\ -m$ respectively
\begin{align}
h(\Delta k_y^+,k_z)&\approx  2m \Delta k_y^+ \tau_x + k_z \tau_y - m \delta \tau_z, \\
h(\Delta k_y^-,k_z)&\approx  -2m \Delta k_y^- \tau_x + k_z \tau_y + m \delta \tau_z, 
\end{align}
where $\Delta k_y^\pm=k_y\mp m$.
Since each of the lowest energy point contributes $\rm{sgn}(\delta)/2$ Chern number, the Chern number $M_c^+=\rm{sgn}(\delta)$ at $R_x=1$ is always non-zero. It is known that non-zero Chern number always leads to the presence of the chiral surface modes at any terminations. In other words, the sign and strength of $\delta$ does not affect the presence of the surface states as illustrated in Fig.~\ref{CaAgAs_surface}(c,d). Thus, even in the absence of the spin-orbital coupling $\delta=0$, the surface states should be present at any termination as shown in Fig.~\ref{CaAgAs_surface}(e). Due to $C_3$ rotation symmetry, the surface states appear at the three planes ($k_x=0,\ k_x=k_y/\sqrt{3},\ k_x=-k_y/\sqrt{3}$). Therefore, the presence of the surface states in the nodal ring stems from the topological phase transition between the two non-trivial topological crystalline insulator phases preserving the reflection symmetry in the $x$ direction, although we note that the presence of the surface states in the three reflection planes does not promise the appearance of the drumhead surface state. The reflection symmetry in the $z$ direction is unable to promise the presence of the stable surface states on (001) surface since the terminations are not located at any reflection center.








\section{space-time inversion symmetry} \label{time-reversal inversion section}

Space-time inversion (\emph{PT}) symmetry, the composite symmetry of time-reversal and inversion, also can quantize the Berry phase when $d\bk$ is integrated along \emph{any} closed loop in any dimensions. Two distinct $PT$ symmetries are distinguished by $PT^2=\pm 1$. As $PT^2=-1$, each band is 2-fold degenerate and the 1d chain preserving this symmetry is always trivial\cite{{Sato_Crystalline_PRB14}}. Here we consider only $PT^2=1$, which can be realized as electron systems without spin-orbital coupling. Since time-reversal and inversion operators both flip $\bk$ to $-\bk$, the composite symmetry operators keep the same $\bk$. Unlike the reflection symmetric systems, the closed integration path can be arbitrarily chosen without breaking \emph{PT} symmetry. 



	\emph{PT} symmetry operator is the combination of a unitary matrix and complex conjugation $PT=V_k \mathcal{K}(\hat{PT}=\sum C_k^\dagger V_k \mathcal{K} C_{k} )$; the unitary matrix $V_k$ in the unit-cell convention might be $\bk$-dependent. The Hamiltonian preserving $PT$ symmetry obeys 
\bee
V_kH^*(k)V_k^*=H(k)
\ee
By assuming the absence of degenerate states, the relation of wavefunctions under \emph{PT} symmetry is given by 
\bee
\ket{\phi _{\bk,j}}=e^{i\beta_{\bk}^j}V_k\ket{\phi _{\bk,j}^*}
\ee
In the unit-cell convention, the Berry phase with a non-contractible integral path can be written as 
\begin{align}
\mathcal{P}=&-i \oint  \sum_{E_j<E_{\textrm{F}} }      \bra{\phi_{\bk,j}} \partial_{\bk} \ket{\phi_{\bk,j}} d\bk  \nonumber \\
=& -i \oint  \sum_{E_j<E_{\textrm{F}} }      \bra{\phi_{\bk,j}^*} V_k^\dagger e^{-i\beta^j_\bk} \partial_{\bk} e^{i\beta^j_{\bk}} V_k \ket{\phi_{\bk,j}^*} d\bk \nonumber \\
=&  \sum_{E_j<E_{\textrm{F}} }  (\beta^j_{+}-\beta^j_{-})-i \oint  \sum_{E_j<E_{\textrm{F}}}      \bra{\phi_{\bk,j}^*} \partial_{\bk}  \ket{\phi_{\bk,j}^*} d\bk \nonumber  \\
& -i \oint  \sum_{E_j<E_{\textrm{F}} }   \bra{\phi_{\bk,j}^*} V_k^\dagger (\partial_{\bk} V_k ) \ket{\phi_{\bk,j}^*} d\bk  \label{PT1}
\end{align}
where $\beta^j_{\mp}$ represent the phases at the beginning and the end of the integration path respectively; hence, the first summation is $2n\pi $, where $n$ is an integer. We use the identity 
\bee
 \bra{\phi _{\bk,j}^*} \partial_{\bk}  \ket{\phi _{\bk,j}^*}= \braket{\partial_{\bk} \phi_{\bk,j}}{\phi_{\bk,j}}=- \bra{\phi_{\bk,j}} \partial_{\bk}  \ket{\phi_{\bk,j}} \label{conjugate}
\ee
If $V_k$ is momentum-independent, $\partial_{\bk} V_k =0$ leads to quantized Berry phase 
\bee
\mathcal{P}= \sum_{E_j<E_{\textrm{F}}}  (\beta^j_{+}-\beta^j_{-})/2  = 0,\ \pi\ (\rm{mod}\ 2\pi)
\ee 
in the unit-cell convention. The requirement of $k$-independent $PT$ is that the inversion image of each atom in the unit cell must be in the same unit cell since time reversal operation does not change atom location. When the center of the unit cell is the inversion center and no atoms are located at the unit cell boundary, $PT$ operator always is $k$-independent. If the reflection image of the atoms in the same unit cell are located at different unit cells, $PT$ operator is $k$-dependent in the unit-cell convention since the situation is similar with reflection symmetry (cf eq.~\ref{k-dependent R}). 
However, in general $V_k$ is momentum-dependent, because the last term in \ref{PT1} might be non-zero and unquantized (see an example in appendix \ref{PT_unit}).

Now we consider the Berry phase in the cell-periodic convention. The space-time inversion operator in this convention is given by 
\bee
PT'=UV_kU \mathcal{K} =e^{-2i\vec{d_m}\cdot \vec{k}}V_0\mathcal{K},
\ee
where $\vec{d_m}$ is the distance between the origin point and the inversion center. (The proof is similar with the reflection operator in the cell-periodic convention in Append.~\ref{Ratomdistance}.)
The Berry phase in the cell-periodic convention is written as
\bee
\mathcal{P}'= \sum_{E_j<E_{\textrm{F}}}  (\beta^j_{+}-\beta^j_{-})/2  -2\pi d_m^\parallel n_0,
\ee 
where $d_m^\parallel$ is the projection of $\vec{d_m}$ along a non-contractable integral direction. For example, if the integral path is from $k_x=0$ to $k_x=2\pi$, $d_m^\parallel$ is given by the $x$-direction projection of $\vec{d_m}$. As the integral path is a closed loop without passing through the whole length of the Brillouin Zone, $d_m^\parallel$ vanishes so that the Berry phase is always quantized. On the other hand, when the unit cell center is chosen at an inversion center, $\vec{d}_m$ vanishes. Furthermore, in the unit-cell convention if $PT$ is momentum-independent, the Berry phases in the two conventions are identical $\mathcal{P}=\mathcal{P}'$ and quantized. 

	To discuss the presence of surface states, we consider $(100)$ surface and the remaining directions are in the periodic boundary condition with momentum $\tilde{k}$. The integral path of the Berry phase in the cell-periodic convention is chosen from $k_x=0$ to $k_x=2\pi$ with fixed $\tilde{k}$.  The center of the Wannier function is given by 
\bee
\bar{x}(\tilde{k})=\frac{a}{2\pi}\mathcal{P}'(\tilde{k}),
\ee
which is identical to Eq.~\ref{center_W}. By following the similar discussion in sec.~\ref{surface_state}, when an inversion center has the same $x$ position of the origin point, which is also the center of the unit cell, and $\mathcal{P}'(\tilde{k})=\pi$, a stable surface state can be present only at the unit cell boundary as a termination, since the termination includes another inversion center at a half lattice constant away from the origin. If some atoms are located at the unit cell boundary, the surface state might be absent since the termination cannot include the inversion center. Furthermore, other terminations, which do not include any inversion center, cannot possess any stable surface states protected by space-time inversion symmetry. 

\subsection{CaP$_3$} \label{CaP3}

When the system preserves only space-time inversion symmetry, a stable surface state can be present \emph{only} at the termination passing through the inversion center. If the inversion centers are away from the termination, the presence of the surface state might not stem from space-time inversion symmetry. Once a surface state appears on the surface not including any inversion center, this surface state must be protected by \emph{other} symmetries or can be adiabatically removed to become the bulk state. 

\begin{figure}[t!]
\begin{center}
\includegraphics[clip,width=0.95\columnwidth]{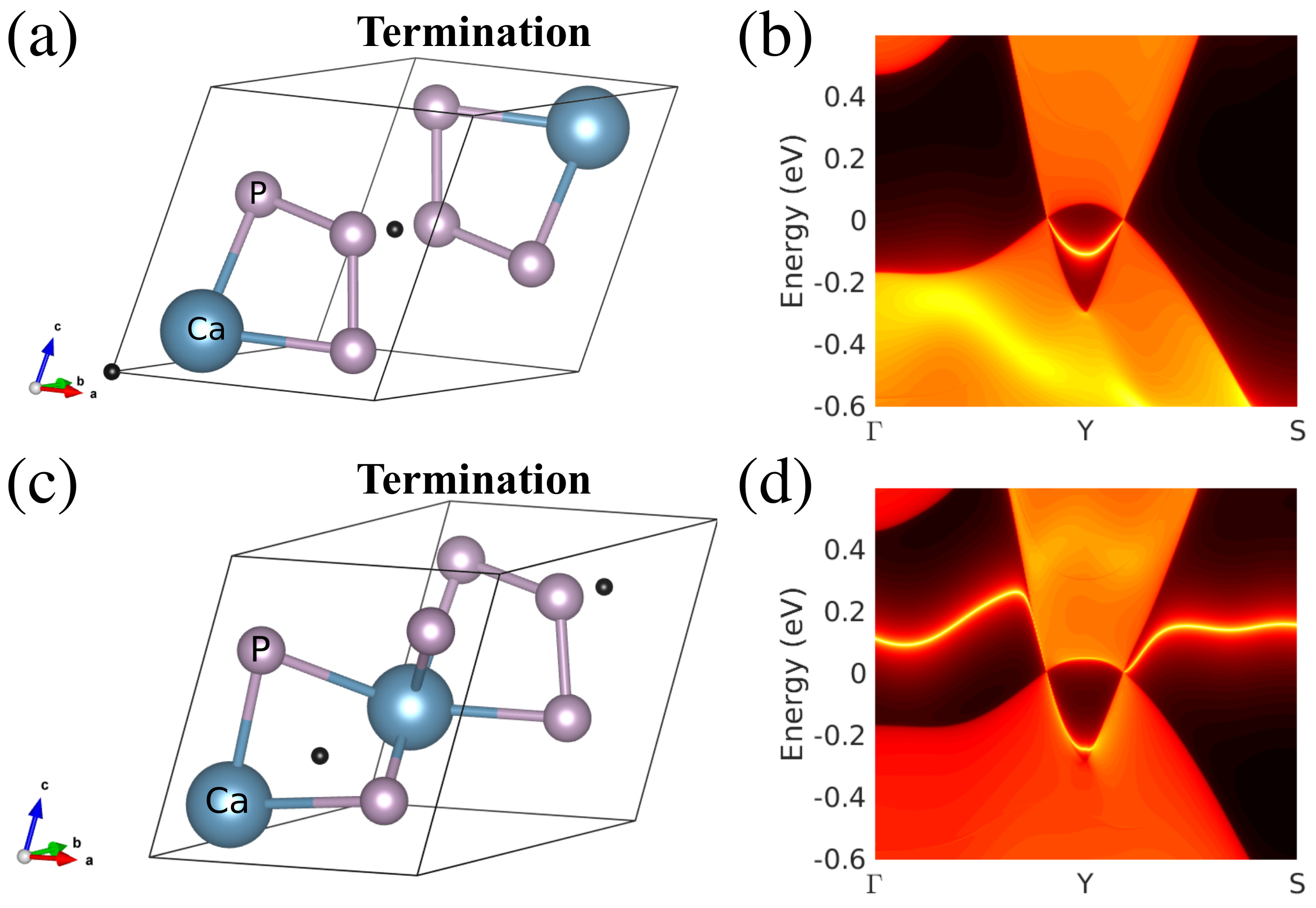}
  \caption{ (a,c) show the crystal structure of CaP$_3$. The black points represent the inversion centers. The terminations on the top with/without the inversion center in panel (a)/(c). (b,d) show for the termination (a,c) respectively on the $(001)$ surface the surface state comes out from the bulk nodal ring in the energy dispersion. Therefore, the surface state still appears even when the termination does not include the inversion center.  } 
  \label{CaP3_structure}
  \end{center}
\end{figure}	


	The following example of CaP$_3$ shows that the presence of the surface state at the termination not including any inversion center stems from time-reversal symmetry. It has been computed in the \emph{ab initio} simulation that CaP$_3$ possesses a 4-fold degenerate nodal ring at $k_z=0$ plane and its surface state connecting the nodal bulk ring appears on the $(001)$ surface passing through the inversion center\cite{PhysRevB.95.045136} as shown in Fig.~\ref{CaP3_structure}(a,b). The space group of the material is $P\bar{1}$ (SG $\#2$), which has only inversion symmetry. Since the system also preserves time-reversal symmetry and spin-$SU(2)$ symmetry, the nodal ring can be block-diagonalized to two spinless nodal rings and these nodal rings are protected by space-time inversion symmetry. Interestingly, we further discover that even if the termination does not include the inversion centers, the (001) surface at any termination possesses the surface states as shown in Fig.~\ref{CaP3_structure}(c,d). In this regard, the presence of the surface states is not related to space-time inversion symmetry. Instead, the non-zero $\bZ_2$ time-reversal topological invariant leads to the presence of the surface state at any termination. 
	
	The low-energy Hamiltonian near the mirror plane $k_z=0$ describing the 4-fold degenerate nodal ring can be written as  
\bee
H(k_\parallel,k_z)=(k_\parallel^2-m^2)\tau_x\otimes \sigma_0 + k_z \tau_y \otimes \sigma_0,  \label{CaP_H}
\ee
where $\sigma_\alpha$ indicates spin degree of freedom while $\tau_\beta$ indicates two different atom locations. 
The system preserves time reversal symmetry 
\bee 
T H({\bf -k})T^{-1}= H({\bf k})
\ee
with $T=\tau_0\otimes \sigma_y \mathcal{K}$ and inversion symmetry with $P=\tau_x\otimes \sigma_0$. The matrix $\tau_x$ represents the exchange of the two atom positions under inversion. When spin-$SU(2)$ symmetry is preserved, the nodal lines at $k_\parallel=m$ are protected by space-time inversion symmetry with $PT^2=\bI$ in the two effective spinless Hamiltonian. 

\begin{figure}[t!]
\begin{center}
\includegraphics[clip,width=0.95\columnwidth]{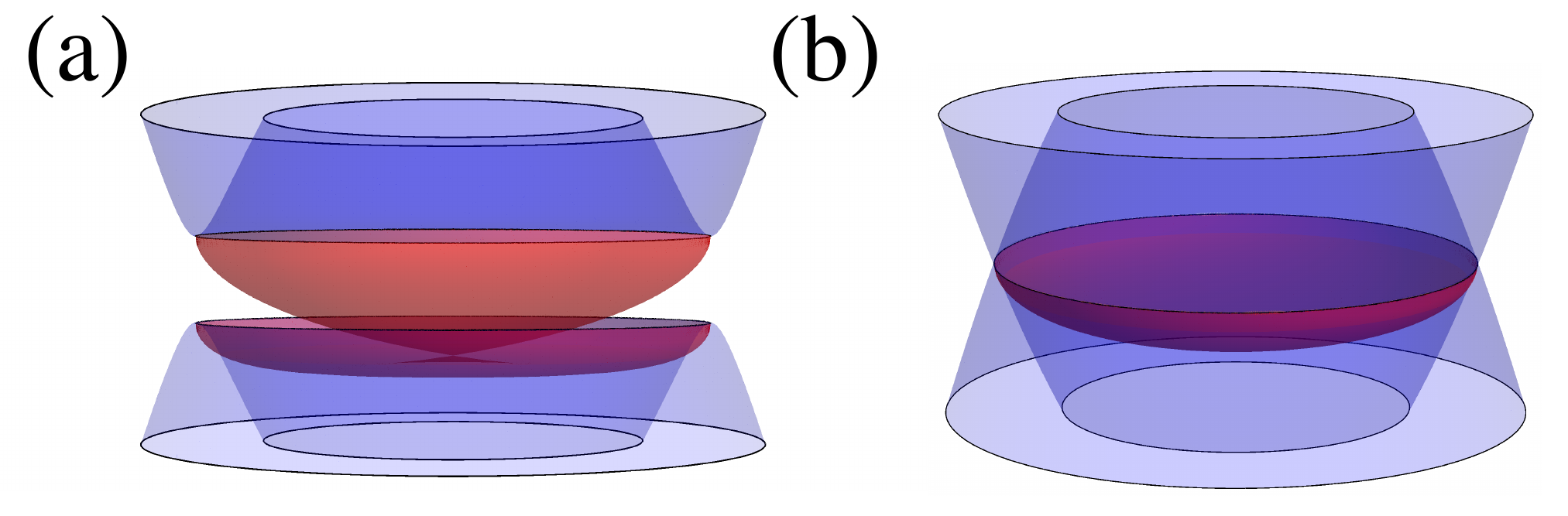}
  \caption{ (a) the surface Dirac cone is present when the nodal ring is gapped due to the non-zero $\bZ_2$ time-reversal invariant. Since the $\bZ_2$ time-reversal invariant is well-defined in the presence of the nodal ring and non-zero, the surface state coexists with the nodal ring as illustrated in (b).   } 
  \label{CaP3_surface}
  \end{center}
\end{figure}

Since inversion symmetry and time-reversal symmetry are preserved, we can compute the $\bZ_2$ time-reversal topological invariant at high symmetry points even if the entire system is a semimetal\cite{PhysRevB.76.045302}. In this low-energy model, we focus on four high symmetric points $(0,0,0),\ (\pi,0,0),\ (0,\pi,0),\ (\pi,\pi,0)$ by assuming no band inversion at the other high symmetric points. The $\bZ_2$ invariant is given by the number occupied kramer's pair in $P=-1$ eigenspace 
\bee
N_{\bZ_2}=\delta_{0,0}\delta_{\pi,0}\delta_{0,\pi}\delta_{\pi,\pi}=-1. 
\ee
Therefore, the nontrivial $Z_2$ invariant leads to the presence of the surface states at any termination since the $Z_2$ time-reversal invariant is not sensitive to the location of the termination. While we relax spin $SU(2)$ symmetry, several candidates, which can destroy the nodal lines, are given by  
\begin{align}
 \tau_z \otimes \sigma_0,\ \tau_z \otimes \sigma_x,\ \tau_z \otimes \sigma_y,\ \tau_z\otimes \sigma_z
\end{align}    
Although $\tau_z \otimes \sigma_0$ preserves spin $SU(2)$ symmetry, $o({\bf k})\tau_z \otimes \sigma_0$ breaks time-reversal symmetry and $e({\bf k})\tau_z \otimes \sigma_0$ breaks inversion symmetry, where $o({\bf k})/e({\bf k})$ is an odd/even function of $k$. Therefore, $\tau_z \otimes \sigma_0$ is forbidden by the symmetries. On the other hand, $o({\bf k})\tau_z\otimes \sigma_j$, which preserves both time-reversal symmetry and inversion symmetry, is allowed to be introduced in the Hamiltonian (\ref{CaP_H}). To completely gap out the nodal lines, symmetry preserving terms $k_x\tau_z \otimes \sigma_x$ and $k_y\tau_z \otimes \sigma_y$ can be added so that the entire system becomes a non-trivial $\bZ_2$ strong time-reversal symmetric topological insulators. As illustrated in Fig.~\ref{CaP3_surface}(a), the surface surface Dirac cone connects the gapped nodal ring in the non-trivial $\bZ_2$ topological insulator. The gapless surface states, which may not stem from the Berry phase quantized by space-time inversion symmetry, can be present at any terminations. Thus, without the spin-orbital coupling, in Fig.~\ref{CaP3_surface}(b) the surface state naturally appears at any termination due to the non-zero $\bZ_2$ time-reversal invariant.

\begin{figure*}[t!]
\begin{center}
\includegraphics[clip,width=1.6\columnwidth]{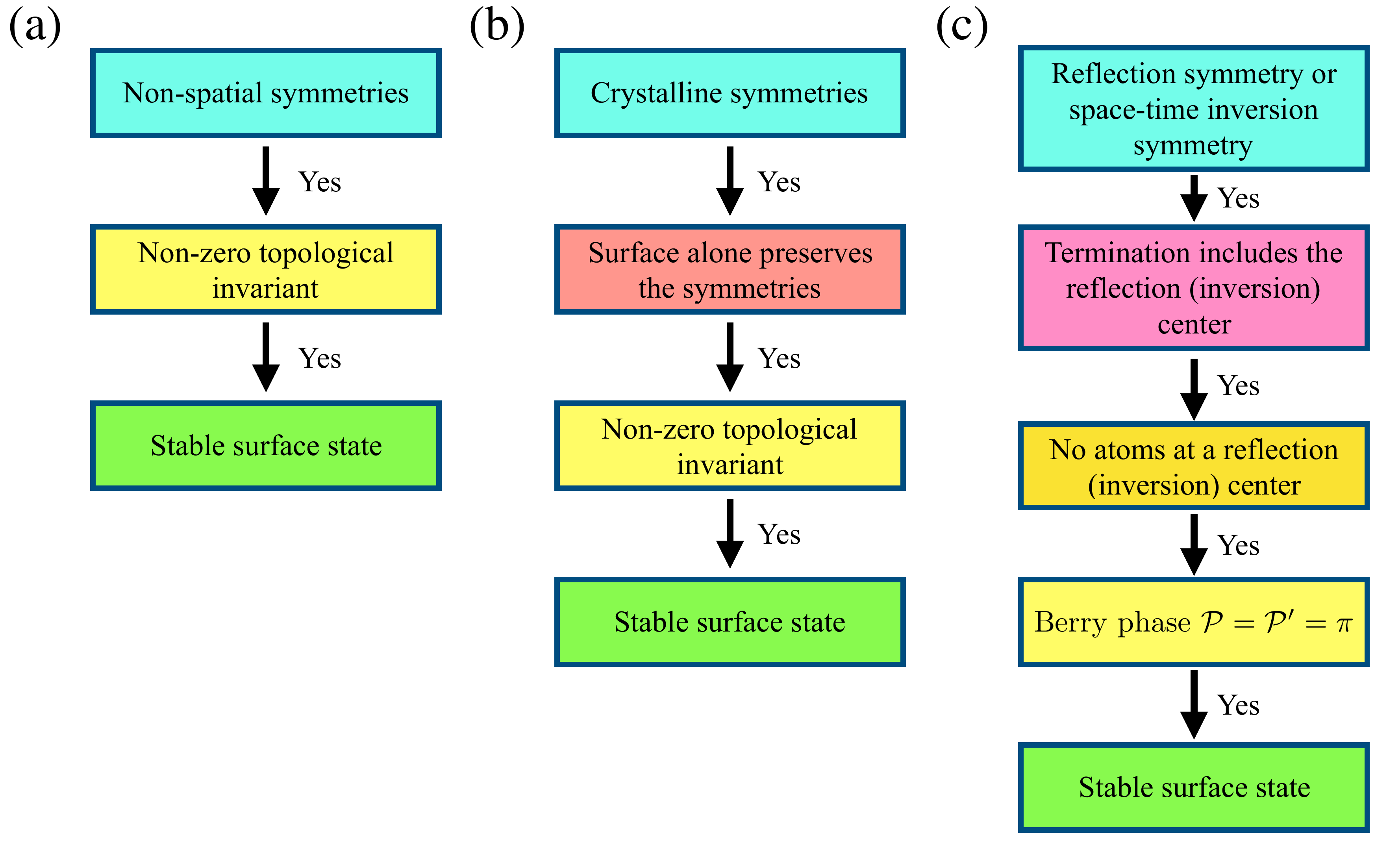}
  \caption{ The criteria leading to the stable surface states for three different topological systems. (a) for the ten-fold classification, once the topological invariant is non-zero, gapless stable surface states are present at any termination on any surface. (b) for topological crystalline insulators and superconductors, the surface alone has to preserve the crystalline symmetries as an additional condition to host gapless stable surface states. (c) for reflection symmetry and space-time inversion symmetry, when the surface alone is not invariant under the symmetry operation. The criterion of the stable surface state is much more complicated.  } 
  \label{surface_state_diagram}
  \end{center}
\end{figure*}

\section{conclusion} \label{conclusion}

	The presence of the surface surface protected by reflection symmetry requires the restrict conditions.  First, the termination, which is the boundary of the unit cell, must be at one reflection center. By moving the spatial origin point to the center of the unit cell, which is another reflection center, the Berry phase in the cell-periodic convention is either $0$ or $\pi$ (mod $2\pi$). These two values correspond the two distinct topological phases. However, the $\pi$-Berry phase is not enough to lead to the presence of the stable surface state at the selected termination. Second, it requires that no atoms can be exactly located at the termination having the reflection center. Since the termination cannot cut any atom into halves, the termination has to move away from the reflection center and then the surface state loses the symmetry protection. When there are no atoms at the reflection center termination, which is the unit cell boundary, the reflection symmetry operator with the reflection center at the unit cell center can be momentum-independent. The Berry phase in the unit-cell convention is quantized and identical to the one in the cell-periodic convention. Finally, under those conditions the $\pi$-Berry phase leads to the presence of the protected surface state at the unit cell boundary. Space-time inversion symmetry shares the same criterion of the stable surface state with reflection symmetry. Hence, this is a completely different type of the stable surface states, since the termination condition plays an important role to determine the presence of the stable surface states. Conventional topological (crystalline) insulators and superconductors have simpler  requirements for the stable surface state as shown in Fig.~\ref{surface_state_diagram}. 
	
	
Since the presence of the stable surface state deeply depends the choice and conditions of the termination, the stability of the drumhead surface states in most of the topological nodal line semimetals have to be reexamined (Mackay-Terrones crystals\cite{PhysRevB.92.045108}, Cu$_3$NPd\cite{PhysRevLett.115.036806,PhysRevLett.115.036807}, CaTe\cite{Du:2017aa}, Ca$_3$P$_2$\cite{nodal_line_Yang}, BaSn$_2$\cite{PhysRevB.93.201114}, Hyperhoneycomb Lattices\cite{PhysRevLett.115.026403}, PbTaSe$_2$\cite{Bian:2016aa}, TlTaSe$_2$\cite{PhysRevB.93.121113}). Although the nodal lines are protected by reflection symmetry or space-time inversion symmetry, the drumhead surface state at the terminations away from the symmetry centers might lose the symmetry protection. The protection might stem from other symmetries.  
	
\section{Acknowledgement}
	
	The authors thank A. Alexandradinata, Ting Cao, Yea-Lee Lee, T.-C. Wei, and Ting Cao for fruitful discussions. C.-K.C. acknowledges support by Microsoft, the Laboratory for Physical Science, and the Strategic Priority
Research Program of the Chinese Academy of Sciences, GrantvNo. XDB28000000.

\appendix

\section{Reflection symmetry}

\subsection{Unquantized Berry phase in unit cell convention}\label{unBerry}
\begin{figure}[t!]
\begin{center}
\includegraphics[clip,width=0.99\columnwidth]{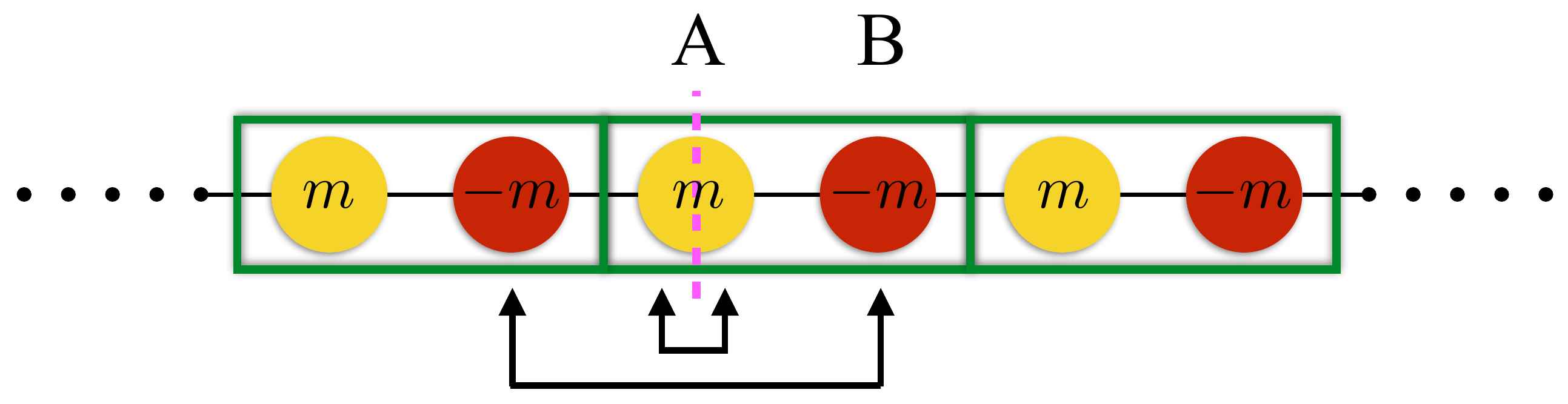}
  \caption{ The simple two-atom model preserving reflection symmetry demonstrates the Berry phase in the unit-cell convention is not quantized. } 
  \label{Example_Appendix_model}
  \end{center}
\end{figure}

We can further show the Berry phase is unqauntized and cannot be connected to the reflection occupation number $n_0^++n_\pi^+$  by using a simple example of a reflection symmetric Hamiltonian in the unit-cell convention. The system can be realized as a 1D spinless chain with the unit cell including two distinct atoms. The tight-binding model is written as  
\bee
\hat{H}_r=\sum_j \Big ( m(A^\dagger_j A_j - B^\dagger_j B_j)+( A^\dagger_j B_{j-1}+ B^\dagger_j A_{j+1}  +  A^\dagger_j B_j + B^\dagger_j A_j ) \Big ).
\ee
Fig.~\ref{Example_Appendix_model} shows the system preserves reflection symmetry and the reflection centers are located at either atom A or atom B. The Hamiltonian in the momentum space is rewritten as 
\bee
H_r(k)=(1+\cos k)\sigma_x + \sin k \sigma_y + m\sigma_z, \label{exampleH}
\ee 
in the basis of $(A_k,B_k)$ for the Pauli matrices $\sigma_i$. The reflection operator (\ref{k-dependent R}) with the reflection center located at atom B is momentum-dependent 
\bee
R_k=
\bma 1 & 0 \\
0 & e^{-ik} \\
\ema. \label{exampleR}
\ee
The Hamiltonian, which obeys $R_kH_r(k)R_k^{-1}=H_r(-k)$ (eq.~\ref{reflection eq}), has two eigenenergies 
\bee
E_{\pm}=\pm \sqrt{ m^2+2+2 \cos k} \label{energy_model}
\ee
We compute the Berry phase of the normalized negative energy state, which is given by as $m>0$
\bee
\ket{\phi_-^{m+}(k)}=\frac{1}{\sqrt{2E_+(E_++m)}}
\bma 
1+\cos k - \sin k i \\
-E_+ -m  \\
\ema
\ee
The wavefunctions at $k, -k$ are connected by the reflection operator
\bee
R_k\ket{\phi_-^{m+}(k)}=\ket{\phi_-^{m+}(-k)}
\ee
Hence, $\alpha_0, \alpha_\pi$ vanishes in Eq.~\ref{Berryphase} so that the Berry phase is given by 
\bee
\mathcal{P}=i \int_{0}^{\pi}      \bra{\phi_-} R_{k}^\dagger (\partial_{k} R_{k} )\ket{\phi_-} dk
\ee
Except for $k=0,\ \pi$, the wavefunction cannot be block diagonalized by the reflection operator; therefore, the wavefunction mixture leads to unqauntized $i \int_{0}^{\pi}      \bra{\phi_-} R_{k}^\dagger (\partial_{k} R_{k} )\ket{\phi_-} dk$. Hence, the unqauntized Berry phase can be directly revealed in its analytic solution
\begin{align}
\mathcal{P}
=&-\frac{1}{2}\bigg (\pi - \frac{2m K(\frac{4}{4+m^2})}{\sqrt{4+m^2}} \bigg ), \label{BerryR}
\end{align}
where $K(a^2)=\int_0^{\pi/2}\frac{d\theta}{\sqrt{1- a^2 \sin^2 \theta}}$ is the incomplete elliptic integral of the first kind. 

	The reflection occupation number can be easily computed by using eq.~\ref{occupation number}. For $m>0$, $n_0^+ + n_\pi^+=1$ since one of the entries in the reflection operator $R_k$ flips the sign. The Berry phase, which is not quantized, is not related with this reflection occupation number. Furthermore, since all of the atoms in this system are located at the reflection centers and the occupied states all are extended, the stable surface states protected reflection symmetry are always absent (sec.~\ref{surface_state}). The odd occupation number $n_0^+ + n_\pi^+$ cannot be an indicator leading to any physical feature for momentum-dependent reflection operator. 


\subsection{Reflection operator in the cell-periodic convention}\label{Ratomdistance}

\begin{figure}[t!]
\begin{center}
\includegraphics[clip,width=0.99\columnwidth]{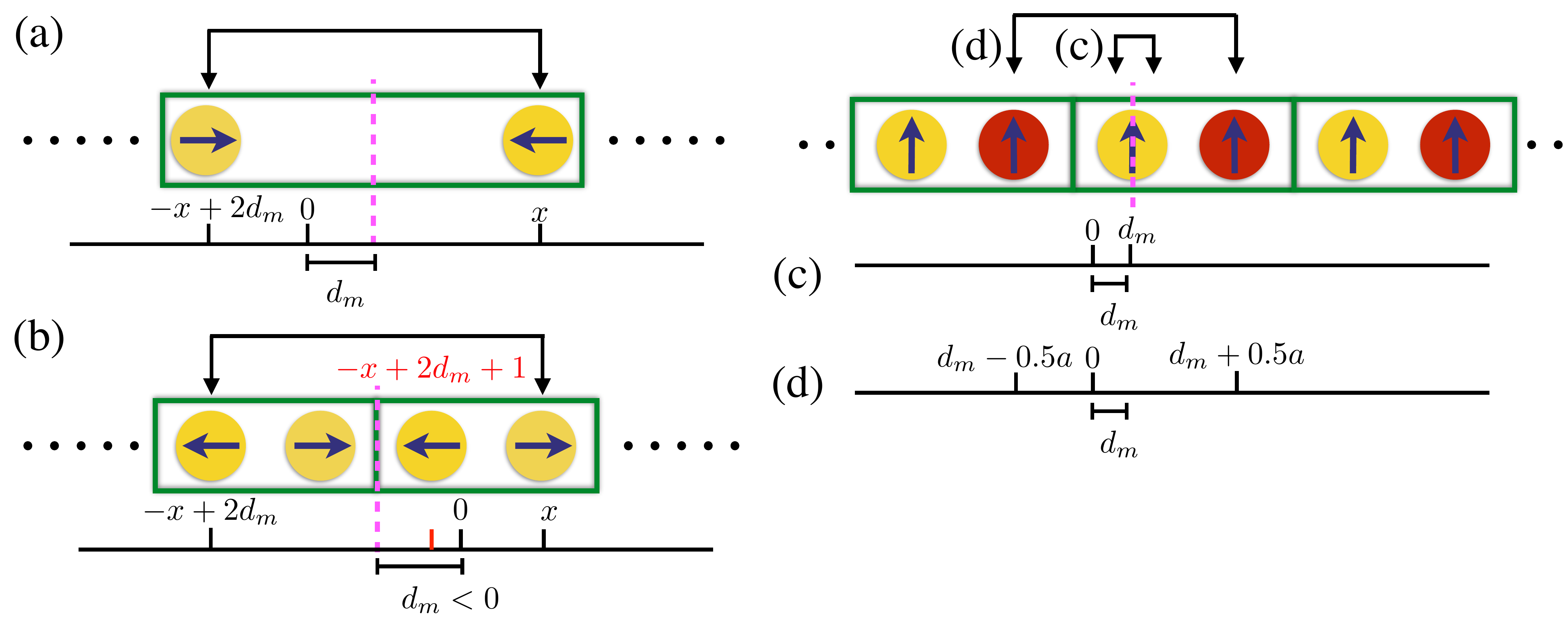}
  \caption{ All possible configurations of the reflection operations (a) an atom $x$ is reflected to another atom ($-x+2 d_m$) in the same unit cell. (b) an atom $x$ is reflected to another atom ($-x+2 d_m$) in the different unit cell. (c) an atom $x=d_m$ is reflected to itself ($x=d_m$) in the same unit cell. The red color indicates the reflection image of the atom at $-x+2d_m+1$ shift back to the same unit cell. (d) an atom $x=d_m\pm 0.5$ is reflected to an identical atom ($x=d_m\mp 0.5$) in the different unit cell.  } 
  \label{reflection_atom_distance_basis}
  \end{center}
\end{figure}

The matrix form of the reflection operator in the cell-periodic convention is simpler than the one in the unit-cell convention. In this section, we show that the reflection operator can be written in the form of a momentum-independent matrix with a momentum-dependent global phase in Eq.~\ref{R cell-periodic}.
The reflection operator in the unit-cell convention is transformed to the cell-periodic convention by Eq.~\ref{two conventions} 
\bee
R_k'=U^\dagger (k) R_k U^\dagger (k)
\ee
To simplify the problem, we set the origin point at the center of the unit cell and $d_m$ is the distance between the reflection center and the origin point. Label $x$ indicates the position of the atom
\bee
x \xrightarrow{R_k} -x+2d_m
\ee
Since there are four possible atom configurations under reflection, we discuss each case individually to find the explicit form of the reflection operator in the cell-periodic convention. 
(a) the reflection operator interchanging two atoms in the same unit cell as illustrated in Fig.~\ref{reflection_atom_distance_basis}(a) is written as 
\bee
R_k=
\bma 
0 & 1 \\
1 & 0
\ema  \otimes S
\ee
The $k$-independent matrix $S$ represents degree of freedom inside the atom, such as spin index, and $S^2=\bI$. The transformation matrix is given by 
\bee
U=\bma  
e^{-i(-x+2d_m)k} & 0 \\
0 & e^{-ixk} 
\ema \otimes \bI, 
\ee
where the positions $x$ and $-x+2d_m$ are the atom and its reflection image in the same unit cell.  
Hence, the reflection operator in the periodic-cell convention is written as 
\bee
R_k'=U^\dagger (k) R_k U^\dagger (k)= e^{2id_mk}\sigma_x \otimes S
\ee 
(b) the reflection of the atom $B_j$ is another atom $A_{-j-1}$ in another unit cell. Since $-x+2d_m$ is the location of another unit cell, the reflection atom has to shift back (+1) to the original cell unit for the cell-periodic convention as illustrated in Fig.~\ref{reflection_atom_distance_basis}(b).
The transformation matrix between the two conventions is based on these two locations 
\bee
U(k)=
\bma 
e^{-i (-x+2d_m +1)k }& 0 \\
0 & e^{-ixk} 
\ema \otimes \bI
\ee 
The reflection operator in unit cell convention can be written in the form of the annihilation and creation operators
\begin{align}
\hat{R}=& \sum_j \big [ B^\dagger_{j} S A_{-j-1} +A_{-j-1}^\dagger S B_{j}  \big ] \\
=& \sum_k (A_{-k}^\dagger\ B_{-k}^\dagger) e^{-ik}\sigma_x \otimes S
\bma 
A_{k} \\
B_{k}
\ema
\end{align}
Therefore, the reflection operator in matrix form in the unit-cell convention is given by 
\bee
R_k = e^{-ik}\sigma_x \otimes S
\ee
The expression of the reflection operator in periodic cell convention can be written as 
\bee
R_k'=U^\dagger(k)R_k U^\dagger(k)=e^{2id_mk}\sigma_x \otimes S
\ee
(c) as an atom is located at the reflection center ($x=d_m$), the reflection of the atom is itself as shown in Fig.~\ref{reflection_atom_distance_basis}(c)
\bee
d_m \xrightarrow{R_k} d_m.
\ee
The reflection operator in unit cell convention clearly is given by $R_k=S$ describing the reflection operation inside the atom; hence, the reflection operator in the periodic-cell convention is given by 
\bee
R_k'=U^\dagger(k)^2=e^{2id_mk}S, 
\ee
where the transformation matrix is $U(k)=e^{-id_mk}\bI$. \\
(d) when an atom is located at another reflection center ($d_m \pm 0.5$), the reflection of the reflection maps to another unit cell as shown in Fig.~\ref{reflection_atom_distance_basis}(d)
\bee
d_m \pm 0.5 \xrightarrow{R_k} d_m \mp 0.5,
\ee
and the reflection image shifts $\pm 1$ lattice constant back to the original atom    
\bee
d_m \mp 0.5 \xrightarrow{\pm 1} d_m \pm 0.5
\ee
The reflection operator in the form of the second quantization is given by 
\bee 
\hat{R}=\sum_{j}C^\dagger_{-j\mp 1} S C_j=\sum_k C_{-k}^\dagger e^{\mp ik} S C_{k}
\ee
Since the transform matrix $U=e^{-i(d_m \pm 0.5)k}$, the reflection operator in periodic convention is in the same expression 
\bee
R_k'=U^\dagger(k) e^{\mp ik} U^\dagger(k)=e^{2id_mk}S, 
\ee
Thus, the reflection operator in the cell-periodic convention can be written in the form of Eq.~\ref{R cell-periodic}
\bee
R_k'=e^{2id_m k} \mathcal{R}_0,
\ee
where $\mathcal{R}_0$ is a momentum-independent matrix and $\mathcal{R}_0^2 =\bI$.

\subsection{Example for the Berry phase in the cell-periodic convention }\label{Berry_atom}

We consider the toy model in the cell-periodic convention and the reflection center is located at the center of the atom as shown in Fig.~\ref{Example_Appendix_model}; hence, the Hamiltonian \eqref{exampleH} in the unit-cell convention is transformed to the cell-periodic convention  
\bee
H(k)=UH_r(k)U^\dagger =m\sigma_z + 2 \cos (k/2) \sigma_x, \label{example cell periodic}
\ee
where $U=\rm{diag}$$(e^{ik/2},1)$, since $e^{ik/2}$ indicates the location of atom A at $x=-a/2$ and $1$ indicates the location of atom B at the origin point. The reflection symmetry operator in this convention is given by 
\bee
R'_k=U^\dagger R_k U^\dagger = \sigma_0
\ee
The additional phase of the reflection operator is absent since the origin point and the reflection center are at the same location. 
The occupied wavefunction with energy $E_-$ (Eq.~\ref{energy_model}) in the unit-cell convention is given by 
(a) $m>0$
\bee
\ket{\phi_-^{m+}}=\frac{1}{\sqrt{2E_+(E_++m)}}
\bma 
1+\cos k - \sin k i \\
-E_+ -m  \\
\ema
\ee
(b) $m<0$
\bee
\ket{\phi_-^{m-}}=\frac{1}{\sqrt{2E_+(E_+-m)}}
\bma 
-E_+ +m  \\
1+\cos k + \sin k i \\
\ema
\ee
In the cell-periodic convention, the wavefunction is transformed to 
\begin{align}
\ket{u_-^{m+}}&=U\ket{\phi_-^{m+}}=\frac{1}{\sqrt{2E_+(E_++m)}}
\bma 
2\cos (k/2)  \\
-E_+ -m  \\
\ema
\end{align}
\begin{align}
\ket{u_-^{m-}}&=U\ket{\phi_-^{m-}}=\frac{e^{ik/2}}{\sqrt{2E_+(E_++m)}}
\bma 
 -E_+ -m \\
2\cos (k/2)   \\
\ema
\end{align}
The Berry phases are given by 
\begin{align}
-i\int_{-\pi}^{\pi} \bra{u_-^{m+}} \partial_k \ket{u_-^{m+}}dk&=0 \ (\rm{mod}\ 2\pi) \\  
-i\int_{-\pi}^{\pi} \bra{u_-^{m-}} \partial_k \ket{u_-^{m-}}dk&=\pi \ (\rm{mod}\ 2\pi) \label{pi inversion}
\end{align}
The Berry phase is quantized as the origin point is the reflection center. However, $\pi$ Berry phase does not lead to the presence of the surface states since atom A is exactly located at the reflection center and all of the occupied states are extended.  


\section{Example: absence of surface states} \label{no_surface}  
	
	We provide a 1D model showing that for the reflection symmetric systems when the termination is not located at the reflection center, the end states are absent after and before the bulk gap closes. The bulk gap closing point in the 1D model can be extended to a 1D nodal ring protected by reflection symmetry in a 3D semimetal. This result leads to the absence of the drumhead surface states in both sides of the nodal line when the termination is not at any reflection center. 
	
\begin{figure}[t!]
\begin{center}
\includegraphics[clip,width=0.95\columnwidth]{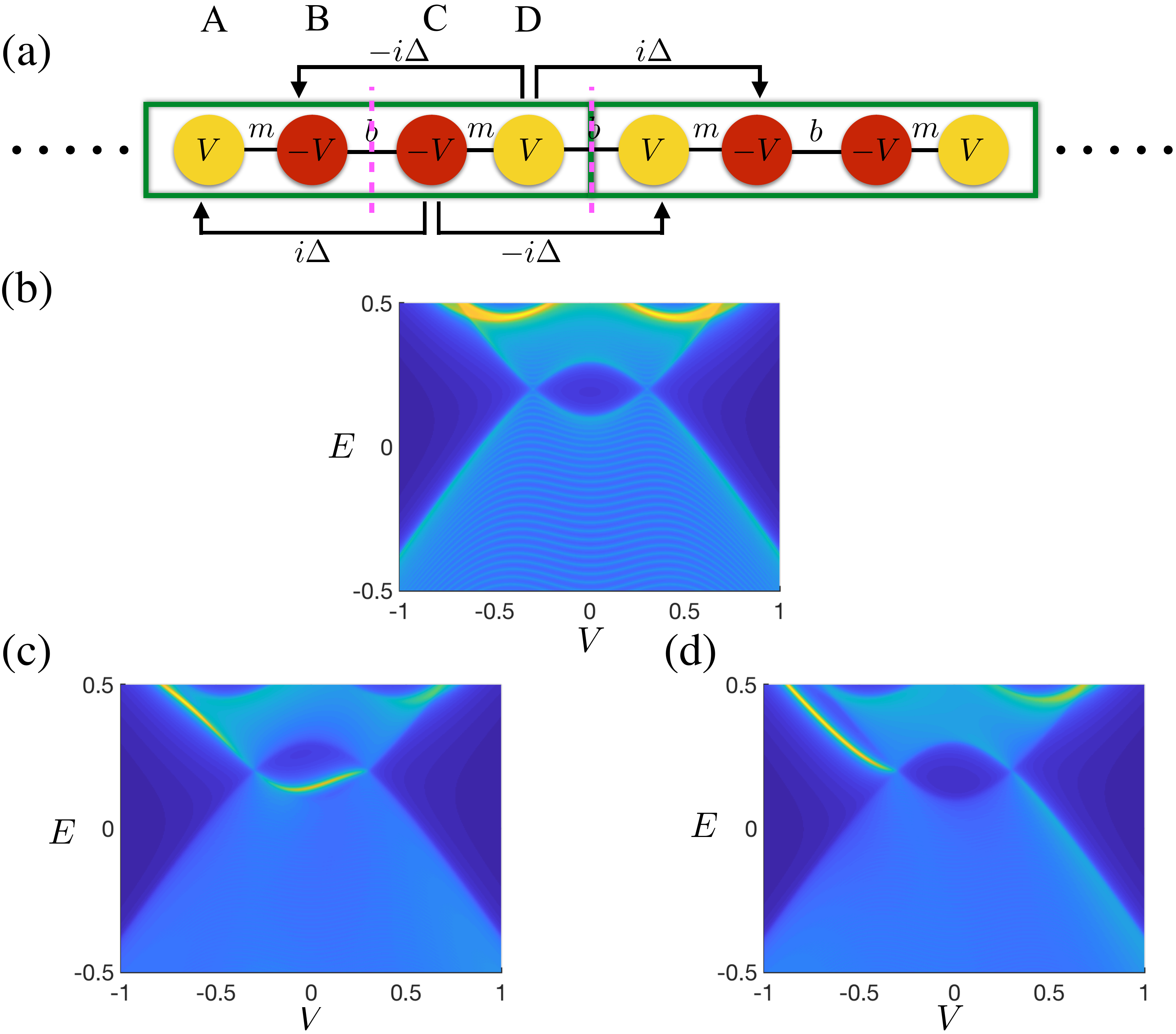}
  \caption{ The four-atom model shows that before and after the bulk gap closing, the surface state is absent when the termination is not at any reflection center. (a) illustrates the unit cell (green box) including four atoms with hoppings and sub-potential. The dashed pink lines represent the reflection centers. (b) the energy dispersion of the bulk states shows the bulk gap closing at $V=\pm 0.3$. (c) the surface state appears for $V<0.3$ at the end of atom A as the termination is at the boundary of the unit cell, which is the reflection center. (d) the surface state is absent for $V>-0.3$ at the end of atom B as the termination is not located at any reflection center. Therefore, the end state is absent near the bulk gap closing point $V=0.3$.  
  } 
  \label{no_surface_state_model}
  \end{center}
\end{figure}

	The model we construct possesses four atoms (A, B, C, D) in the unit cell. We choose the origin point at the center of the unit cell and atom A, B, C, D are located at $x_A=-3a/8,\ x_B=-a/8,\ x_C=a/8,\ x_D=3a/8$ respectively. We build a spinless Hamiltonian in the basis $\Phi_j=(A_j\ B_j\ C_j\ D_j)^T$
\begin{align}
\hat{H}=&\sum_j \Bigg ( \Phi_j^\dagger 
\bma 
V & m & i\Delta & 0 \\
m & -V & b & -i\Delta \\
-i\Delta & b & -V & m \\
0 & i\Delta & m * V \\
\ema
\Phi_j \nonumber \\
&+\bigg(  \Phi_{j+1}^\dagger 
\bma 
c/2 & 0 & -i\Delta & b \\
0 & c/2 & 0 & i\Delta \\
0 & 0 & c/2 & 0 \\
0 & 0 &0 & c/2\\
\ema
\Phi_j + h.c. \bigg ) \Bigg )
\end{align}	
The hopping terms and the sublattice potential are illustrated in Fig.~\ref{no_surface_state_model}(a). The Hamiltonian in momentum space in the unit-cell convention can be written as  
\begin{small}
\begin{align}
&H(k)= \nonumber \\
&\bma 
V+c*\cos k & m & i \Delta (1-e^{-ik}) & b e^{-ik} \\
m & -V + c*\cos k & b & -i{\Delta}(1-e^{-ik}) \\
-i{\Delta}(1-e^{ik}) & b & -V+c*\cos k & m \\
b e^{ik} & i \Delta (1-e^{ik}) & m & V+ c*\cos k 
\ema \label{H_surf}
\end{align}
\end{small}
Since the reflection operation with the reflection center located at the unit cell center exchanges atom A, D and atom B, C, the reflection symmetry operator is given by 
\bee
R=\bma 
0 & 0 & 0 & 1 \\
0 & 0 & 1 & 0 \\
0 & 1 & 0 & 0 \\
1 & 0 & 0 & 0
\ema. 
\ee
The Hamiltonian preserves the reflection symmetry by obeying $R^{-1}H(-k)R=H(k)$. We choose the values of the parameters $m=0.4,\ c=0.2,\ \Delta=0.25,\ b=0.5$ and the sublattice potential $V$ can be tuned. The bulk energy dispersion in Fig.~\ref{no_surface_state_model}(b) shows the bulk gap closing at $V=\pm 0.3=\pm \sqrt{b^2-m^2}$.

According to the locations of the atoms, the unitary matrix transforms the system from the unit-cell convention to the cell-periodic convention can be written in diagonal matrix form $U=\rm{diag}$$(e^{i3k/8},e^{ik/8},e^{-ik/8},e^{-3ik/8})$. To compute the Berry phase in the cell-periodic convention, we first find the two occupied states for the Hamiltonian (\ref{H_surf}) in the unit-cell convention and transform the wavefunctions to the cell-periodic convention. Using the numerical method\cite{kingSmithPRB93b}, we obtain that the Berry phase in the cell-periodic convention is $\pi$ for $-0.3<V<0.3$ and $0$ elsewhere. The termination is at the unit cell boundary, which is a reflection center. As shown in Fig.~\ref{no_surface_state_model}(c), the end state appears at $-0.3<V<0.3$ because $\pi$ Berry phase indicates that a stable state is locked at the boundary of the unit cell, which is the reflection center. The appearance of the end state as $V<-0.3$ is rather accident, since the zero-value Berry does not provide any protection. 

	Now we consider the termination is in the middle of atom A and B, which is not a reflection center, and redefine the origin point in the middle of atom C and D. The Berry phase in periodic cell convention is given by $\pi$ for $0.3<|V|$ and $0$ elsewhere by using Eq.~\ref{Berryphase II}. However, $\pi$ Berry phase cannot lead to the presence of the stable end state since the termination, which is not the reflection center, cannot lock one state. Hence, Fig.~\ref{no_surface_state_model}(d) shows the absence of the end state for $-0.3<V$; the end state can be absent after and before the bulk gap closing point $V=0.3$. Thus, this model can be extended to the reflection symmetric nodal ring semimetals and the drumhead surface is absent inside and outside the nodal ring.  


\section{Space-time inversion symmetry}

\subsection{Unquantized Berry phase in the two conventions with $PT'^2=\bI$} \label{PT_unit}
To explicitly show the unqauntized Berry phase in the unit-cell convention when the space-time inversion symmetry operator $PT$ is momentum dependent, we adopt the Hamiltonian (\ref{exampleH}). The Hamiltonian preserving $PT$ symmetry obeys $PT H_r(k) (PT)^{-1}=H_r(k)$, where the symmetry operator 
\bee 
PT=V_k \mathcal{K}=
\bma 
1 & 0 \\
0 & e^{ik} \\
\ema \mathcal{K}
\ee   
satisfies $PT^2=\bI$. Since the Hamiltonian is identical to the reflection case, the Berry phase is identical to Eq.~\ref{BerryR}. Hence, again the Berry phase is unqauntized. 

	The Hamiltonian (\ref{exampleH}) in the unit-cell convention can be transformed to the expression (\ref{example cell periodic}) in the cell-periodic convention by using the unitary matrix $U=\rm{diag}$$(e^{ik/2},1)$ as the location of atom B is the spatial origin. The space-time inversion operator in this convention is given by 
\bee
PT'=UV_kU\mathcal{K}=e^{ik} \sigma_0 \mathcal{K}
\ee	
The Hamiltonian (\ref{example cell periodic}) obeys space-time inversion symmetry equation $H'^*(k)=H(k)$. Since the Berry phase has been calculated in appendix \ref{Berry_atom}, eq.~\ref{pi inversion} gives Berry phase $\mathcal{P}'=\pi$ for $m<0$. Again, since all of the atoms are located at the inversion centers and all of the occupied states are extended. Therefore, it is not possible to possess an end state protected by space-time inversion symmetry.

\bibliographystyle{apsrev4-1}

\bibliography{TOPO3_v14}

\end{document}